\documentclass[aps,prd,a4paper,superscriptaddress,nofootinbib,
showkeys,amsfonts,amssymb,amsmath,11pt]{revtex4}
\usepackage{amssymb,latexsym}
\usepackage{amsmath, amsthm}
\usepackage{amscd}
\usepackage{amsthm}
\usepackage{times}
\usepackage{epsfig}
\usepackage{psfrag}
\usepackage{graphicx}
\usepackage{amssymb,latexsym}

\begin{document}
\title{Spacetime Singularities}
\author{Pankaj S. Joshi} \email{psj@tifr.res.in}
\affiliation{Tata Institute of Fundamental Research, Homi 
Bhabha road, Colaba, Mumbai 400005, India}
\def\m{$M$}
\def\mb{$\overline M$}
\def\i{$\cal I^+$}
\def\s{spacetime}
\def\p{singularity}
\def\t{\triangle}

\begin{abstract} 
We present here an overview of our basic understanding and 
recent developments on spacetime singularities in the Einstein 
theory of gravity. Several issues related to physical 
significance and implications of singularities are discussed. 
The nature and existence of singularities are considered 
which indicate the formation of super ultra-dense regions in 
the universe as predicted by the general theory of
relativity. Such singularities develop during the gravitational collapse 
of massive stars and in cosmology at the origin of the universe.
Possible astrophysical implications of the occurrence of 
singularities in the spacetime universe are indicated. We  
discuss in some detail the profound and key fundamental 
issues that the singularities give rise to, such as the  
cosmic censorship and predictability in the universe, 
naked singularities in gravitational collapse and their 
relevance in black hole physics today, and their astrophysical 
implications in modern relativistic astrophysics and 
cosmology.

\end{abstract}
\maketitle

\section{Introduction}
After Einstein proposed the general theory of 
relativity describing the gravitational force in terms 
of the spacetime curvatures, the proposed 
field equations related the spacetime geometry to the  
matter content of the universe. The early solutions found 
for these equations were the Schwarzschild metric and
the Friedmann models. While the first represented 
the gravitational field around an isolated body such as a 
spherically symmetric star, the later solutions described
the geometry of the universe.  
Both these models contained a spacetime singularity 
where the curvatures as well as the matter and energy densities 
were diverging getting arbitrarily high, and the physical 
description would then break down. In the Schwarzschild 
solution such a singularity was present at the center of 
symmetry $r=0$ whereas for the Friedmann 
models it was found at the epoch $t=0$ which is beginning 
of the universe and origin of time where the scale factor 
for the universe vanishes and all objects are crushed to a 
zero volume due to infinite gravitational tidal forces.

Even though the physical problem posed by the existence 
of such a strong curvature singularity was realized immediately 
in these solutions, which turned out to have several important 
implications towards the experimental verification of the
general relativity theory, initially this phenomenon was not 
taken seriously. It was generally thought that the existence 
of such a singularity must be a consequence of the very 
high degree of symmetry imposed on the 
spacetime while deriving and obtaining these solutions. 
Subsequently, the distinction between a genuine spacetime 
singularity and a mere coordinate singularity became
clear and it was realized that the singularity at $r=2m$ 
in the Schwarzschild spacetime was only a coordinate singularity 
which could be removed 
by a suitable coordinate transformation. It was clear, however, 
that the genuine curvature singularity at
$r=0$ cannot be removed by any coordinate transformations. 
The  hope was then that when more general
solutions to the field equations are considered with a 
lesser degree of symmetry, such singularities will be avoided.

This issue was sorted out when a detailed study of 
the structure of a general spacetime and the associated 
problem of singularities was taken up by Hawking, Penrose, 
and Geroch (see for example, 
Hawking and Ellis, 1973, 
and references therein). 
It was shown by this work that a spacetime will admit
singularities within a rather general framework provided it 
satisfies certain reasonable physical assumptions such as 
the positivity of energy, a suitable causality assumption and 
a condition implying strong gravitational fields, 
such as the existence of trapped surfaces.
It thus followed that the spacetime singularities form 
a general feature of the relativity theory. In fact, these 
considerations ensure the existence of singularities in other 
theories of gravity also which are based on a spacetime 
manifold framework and that satisfy the general conditions 
such as those stated above.

Therefore the scenario that emerges is the following: 
Essentially for all classical spacetime theories of gravitation, 
the occurrence of singularities form an inevitable and integral
part of the description of the physical reality. In the vicinity
of such a singularity, typically the energy densities, spacetime 
curvatures, and all other physical quantities would blow up,
thus indicating the occurrence of super ultra-dense regions 
in the universe. The behaviour of such regions may not be 
governed by the classical theory itself, which may breakdown 
having predicted the existence of the singularities, and 
a quantum gravitation theory would be the most likely 
description of the phenomena created by such singularities.

Further to the general relativity theory in 1915, the 
gravitation physics was a relatively quiet 
field with few developments till about 1950s. However, 
the 1960s saw the emergence of new observations in high energy 
astrophysics, such as quasars and high energy phenomena at 
the center of galaxies such as extremely energetic jets. 
These observations, together with important theoretical 
developments such as studying the global structure of spacetimes 
and singularities, led to important results in black 
hole physics and relativistic astrophysics and cosmology.

Our purpose here is to review some of these rather interesting
as well as intriguing developments, in a somewhat pedagogic and 
elementary fashion providing a broad perspective. Specifically, 
we would like to highlight here several recent issues and 
challenges that have emerged related to spacetime singularities, 
which appear to have a considerable physical significance and 
may have interesting astrophysical implications.  
We take up and consider here important topics such as, 
what is meant by a singular \s\  and specify the notion of 
a \p. It turns out that it is the notion of  geodesic incompleteness 
that characterizes a singularity in an effective manner for 
a \s\  and enables their existence to be proved by means of 
certain general enough theorems. 
We highlight here several recent developments which 
deal with certain exciting current issues related to spacetime 
singularities on which research in gravitation and cosmology 
is happening today. These include the work on final endstates 
of gravitational collapse, cosmic censorship, 
and black holes and naked singularities. Related major 
cosmic conundrums such as the issue of predictability in 
the universe are discussed, and observational implications 
of naked singularities are indicated.

\section{What is a singularity?}
In the Einstein theory of gravitation, the universe is
modeled as a spacetime with a mathematical structure of 
a four dimensional differentiable manifold. In that case, 
locally the spacetime is always flat in a sufficiently small
region around any point, but on a larger scale this need not be 
the case and it can have a rich and varied structure. An 
example of a differentiable manifold is a sphere, which is 
flat enough in the vicinity of any single point on its surface, 
but has a global curvature.
For a more detailed discussion on spacetime manifolds and their
key role in general relativity, we refer to
(Wald 1984).

When should we say that a \s\  universe $(M,g)$,
which is a differentiable manifold with a Lorentzian metric, 
has become singular? What we need is a specification and a 
specific criterion for the existence of a singularity for 
any given universe model in general relativity.

As stated above, several examples of
singular behaviour in \s\  models of general relativity 
are known. Important exact solutions such as the Friedmann$-
$Robertson$-$Walker (FRW) cosmological models and the 
Schwarzschild \s\  contain a \p\ where the energy 
density or curvatures diverge strongly and the usual 
description of the \s\  breaks down. In the Schwarzschild \s\  
there is an essential curvature singularity at $r=0$ in that 
along any nonspacelike trajectory falling into
the singularity, the Kretschmann scalar 
$\alpha=R^{ijkl}R_{ijkl}\to \infty$. Also, all future directed 
nonspacelike geodesics which enter the event horizon at $r=2m$ 
must fall into this curvature singularity within a finite 
value of the proper time, or finite value of the affine 
parameter, so all such curves are future geodesically incomplete.
Similarly, for FRW models, if $\rho+3p>0$ at all times, 
where $\rho$ is the total energy density and $p$ the pressure, 
there is a singularity at $t=0$ which could be identified
as the origin of the universe. Then along all the past directed 
trajectories meeting this singularity, $\rho\to \infty$ 
and the curvature scalar $R=R_{ij}R^{ij}\to \infty$. Again, 
all the past directed nonspacelike geodesics are incomplete. 
This essential singularity at $t=0$ cannot be transformed
away by any coordinate transformations. Similar behaviour was 
generalized to the class of spatially homogeneous cosmological 
models by 
Ellis and King (1974) 
which satisfy the positivity of energy conditions.

Such singularities, where the curvature scalars and densities
diverge imply a genuine \s\  pathology where the usual laws of physics 
break down. The existence of the geodesic incompleteness in these 
spacetimes imply, for example, that a timelike observer suddenly 
disappears from the \s\  after a finite amount of proper time.
Of course, singular behaviour can also occur without bad behaviour of 
curvature. For example, consider the Minkowski \s\  with a point 
deleted. Then there will be timelike geodesics running into the hole 
which will be future incomplete. Clearly this is an
artificial situation one would like to rule out by requiring
that the \s\  is {\it inextendible}, that is, it cannot be 
isometrically embedded into another larger  \s\ 
as a proper subset.
But one could give a non-trivial example of the singular 
behaviour where a conical singularity exists
(see e.g. Ellis and Schmidt, 1977). 
Here \s\  is inextendible but curvature components do not 
diverge near the singularity, as in a Weyl type of 
solution. The metric is given by,
$$ ds^2= -dt^2+dr^2+r^2(d\theta^2+\sin^2\theta\, d\phi^2)$$
with coordinates given by $-\infty<t<\infty, 0<r<\infty,
0<\theta<\pi$ with $0<\phi<a$, with $\phi=0$ and $\phi=a$ 
identified and $a\ne 2\pi$. There is a conical singularity at 
$r=0$ through which the \s\  cannot be extended and the singular 
boundary is related to the timelike two-plane $r=0$ 
of the Minkowski \s .

An important question then is, whether
such singularities develop even when a general model 
is considered, and if so under what conditions. To answer this, 
it is first necessary to characterize precisely 
what one means by a \s\  \p for a general \s. Then it is seen 
that singularities must exist for a very wide class of spacetimes 
under reasonable general set of conditions. Such singularities 
may develop as the end state of gravitational collapse of a 
massive star, or in the cosmological situations such as 
the origin of the universe.

The first point to note here is by very definition,
the metric tensor has to be well defined at all the regular points 
of the \s. This is no longer true at a \s\  \p\ such as 
those discussed above and a \p\ cannot be regarded as a regular 
point of \s\  but is a boundary point attached to \m. This 
causes difficulty when one tries to characterize a 
\p\ by the criterion that the curvatures must blow up 
near the \p. The trouble is, since the singularity is not 
part of the \s , it is not possible to define its neighborhood 
in the usual sense to discuss the behaviour of curvature 
quantities in that region.
One may try to characterize a \p\ in terms of 
the divergence of components of the Riemann curvature tensor 
along non-spacelike trajectories. Then the trouble 
is, the behaviour of such components will in general change 
with the frames used so this is not of much help. One can 
try the curvature scalars or the scalar polynomials in 
the metric and the Riemann tensor and require them to achieve 
unboundedly large values. This is encountered in Schwarzschild 
and the Friedmann models. But it is possible that 
such a divergence occurs only at infinity for a given 
nonspacelike curve. In general it looks reasonable
to demand that some sort of curvature divergence must take 
place along the nonspacelike curves which encounter a \p.
However, a general characterization of \p\ in terms of
the curvature divergence runs into various difficulties.

Considering these and similar situations, the occurrence 
of nonspacelike geodesic incompleteness is generally agreed 
upon as the criterion for the existence of a \p\ for a \s . 
This may not cover all types of singular behaviours possible
but it is clear that if a \s\  manifold contains incomplete 
nonspacelike geodesics, there is a definite singular behaviour 
present, as a timelike observer or a photon suddenly disappears 
from the \s\  after a finite amount of proper time or after 
a finite value of the affine parameter. The singularity
theorems which result from an analysis of gravitational 
focusing and global properties of a \s\  prove this incompleteness 
property for a wide class of spacetimes under a set of 
rather general conditions.

Physically, a singularity in any physics theory
typically means that the theory breaks down either in the 
vicinity or at the singularity. This means that a broader
and more comprehensive theory is needed, demanding a revision
of the given theory. A similar reasoning apply to spacetime 
singularities, which may be taken to imply that a quantum
gravity description is warranted in those regions of the
universe, rather than using merely a classical 
framework.

\section {Gravitational focusing}
The simple model solutions such as the Schwarzschild 
and the FRW universes give very useful indications on what is
possible in general relativity, as opposed to the Newtonian
gravity. In particular, these solutions told us on important 
indicators on the existence and nature of the spacetime 
singularities.

The key to occurrence of singularities in these solutions
is really the gravitational focusing that the matter fields and 
spacetime curvature
causes in the congruences of null and timelike curves, which
represent the light paths and the material particle trajectories
in any given spacetime universe. It would then be important to
know how general and generic such a feature and property is
for a general spacetime.

The matter fields with positive energy 
density which create the curvature in spacetime affect the 
causality relations in a \s\  and create focusing in the 
families of nonspacelike trajectories. The phenomenon 
that occurs here is matter focuses the nonspacelike geodesics 
of the \s\  into pairs of {\it focal points} or {\it conjugate
points}. The property of conjugate points is that if $p, q$ 
are two conjugate points along a nonspacelike geodesic, 
then $p$ and $q$ must be timelike related. Now, there are 
three-dimensional null hypersurfaces such as the boundary of 
the future of an event such that no two points of such a hypersurface 
can be joined by a timelike curve. Thus, the null geodesic 
generators of such surfaces cannot contain conjugate points
and they must leave the hypersurface before encountering any 
conjugate point. This puts strong limits on such null 
surfaces and the singularity theorems result
from an analysis of such limits.

If we consider a congruence of timelike geodesics in the \s , 
this is a family of curves and through each event there passes 
precisely one timelike geodesic trajectory. Choosing the curves 
to be smooth, this defines a smooth timelike vector field on 
the \s . 
The rate of change of volume expansion for a given 
congruence of timelike geodesics can be written as
$${d\theta\over d\tau}=-R_{lk}V^l V^k-\textstyle{1\over3}\theta^2-2\sigma^2
+2\omega^2$$
where, for a given congruence of timelike (or null) geodesics, 
the quantities $\theta$, $\sigma$ and $\omega$ are 
{\it expansion}, {\it shear}, and {\it rotation} tensors 
respectively.
The above equation is called the {\it Raychaudhuri equation} 
(Raychaudhuri, 1955), 
which describes the rate of change of the volume expansion 
as one moves along the timelike geodesic curves in the congruence.

We note that the second and third term on the right-hand side 
involving $\theta$ and
$\sigma$ are positive always. For the term $R_{ij}V^iV^j$, 
by Einstein equations this can be written as
$$R_{ij}V^i V^j= 8\pi[T_{ij}V^i V^j +\textstyle{1\over2}T]$$
The term  $T_{ij}V^i V^j$ above represents the energy density 
measured by a timelike observer with unit tangent $V^i$, 
which is the four-velocity of the observer. For all reasonable 
classical physical fields this energy density is generally 
taken as non-negative and we may assume for all timelike 
vectors $V^i$,
$$T_{ij}V^i V^j\ge 0$$ 
Such an assumption is called the {\it weak energy condition}.
It is also considered reasonable that the 
matter stresses will not be so large as to make the 
right-hand side of the equation above negative. This will 
be satisfied when the following is satisfied,
$ T_{ij}V^i V^j\ge -\textstyle{1\over2}T$.
Such an assumption is called the {\it strong energy condition} 
and it implies that for all timelike vectors $V^i$,
$R_{ij}V^i V^j\ge 0$.
By continuity it can be argued that the same will then 
hold for all null vectors as well.
Both the strong and weak energy conditions will be valid for well-known
forms of matter such as the perfect fluid provided the energy density
$\rho$ is non-negative and there are no large negative pressures which are
bigger or comparable to $\rho$.

With the strong energy condition, the Raychaudhuri equation
implies that the effect of matter on \s\  curvature causes a focusing effect
in the congruence of timelike geodesics due to gravitational attraction.
This causes the neighbouring geodesics in the congruence to cross
each other to give rise to caustics or conjugate points. Such a
separation between nearby timelike geodesics is governed by 
the geodesic deviation equation,
$$ D^2Z^j= {-R^j}_{kil}V^k Z^i V^l$$
where $Z^i$ is the separation vector between nearby geodesics of the 
congruence. Solutions of the above equation are called the {\it Jacobi fields}
along a given timelike geodesic.

Suppose now $\gamma$ is a timelike geodesic,
then two points $p$ and $q$ along $\gamma$ are called {\it conjugate points}
if there exists a Jacobi field along $\gamma$ which is not identically zero
but vanishes at $p$ and $q$. From the Raychaudhuri equation
given above it is clear that the occurrence of conjugate points along a timelike
geodesic is closely related to the behaviour of the expansion parameter $\theta$
of the congruence. In fact, it can be shown that the necessary and sufficient
condition for a point $q$ to be conjugate to $p$ is that for the congruence
of timelike geodesics emerging from $p$, we must have $\theta\to -\infty$
at $q$ (see for example, 
Hawking and Ellis, 1973). 
The conjugate points along null geodesics are also 
similarly defined.


\section{Geodesic incompleteness}
It was widely believed that for more general
solutions of the Einstein equations which incorporate several
other physical features and not necessarily symmetric,
the existence of singularities would be avoided. Further
investigations, however, showed that singularities in the
form of geodesic incompleteness do exist for general
spacetimes. These results used the gravitational focusing
considerations mentioned above and global properties
of a general spacetime.

The behaviour of the expansion parameter $\theta$ is governed by the
Raychaudhuri equation as pointed out above.
Consider for example the situation
when the \s\  satisfies the strong
energy condition and the congruence of timelike
geodesics is hypersurface orthogonal. Then $\omega_{ij}=0$ and the
corresponding term $\omega^2$ vanishes. Then, the expression
for the covariant derivative of $\omega_{ij}$ implies that it must vanish
for all future times as well. It follows that we must have then
$ {d\theta\over d\tau}\le -{\theta^2\over3},$
which means that the volume expansion parameter must be necessarily
decreasing along the timelike geodesics. If $\theta_0$ denotes the initial
expansion then the above can be integrated as $\theta^{-1}\ge
\theta_0^{-1}+\tau/3$. It is clear from this that if the congruence is
initially converging and $\theta_0$ is negative then $\theta\to -\infty$
within a proper time distance $\tau\le 3/\mid \theta_0\mid$.

It then follows that if \m\  is a \s\  satisfying the strong energy condition
and $S$ is a spacelike hypersurface with $\theta<0$ at $p\in S$, then if $\gamma$
is a timelike geodesic of the congruence orthogonal to $S$ passing through
$p$ then there exists a point $q$ conjugate to $S$ along $\gamma$ within a
proper time distance $\tau\le 3/\mid \theta\mid$. This is provided
$\gamma$ can be extended to that value of the proper time.

The basic implication of the above results is that once a convergence
occurs in a congruence of timelike geodesics, the conjugate points or the
caustics must develop in the \s . These can be interpreted as the
singularities of the congruence. Such singularities could occur even in
Minkowski \s\  and similar other perfectly regular \s  s. However, when combined
with certain causal structure properties of \s\, the
results above imply the existence of singularities in the form of
geodesic incompleteness.
One could similarly discuss the gravitational
focusing effect for the congruence of null geodesics or for null geodesics
orthogonal to a spacelike two-surface.

There are several singularity theorems available which establish the nonspacelike
geodesic incompleteness for a \s\  under different sets of conditions and
applicable to different physical situations. However, the most general of
these is the Hawking$-$Penrose theorem
(Hawking and Penrose, 1970),
which is applicable in both the collapse situation and 
cosmological scenario. The
main idea of the proof of such a theorem is the following. Using the causal
structure analysis it is shown that there must be maximal length timelike
curves between certain pairs of events in the \s . Now, a causal
geodesic which is both future and past complete must contain pairs of
conjugate points if \m\  satisfies an energy
condition. This is then used to
draw the necessary contradiction to show that \m\  must be
non-spacelike geodesically
incomplete.

{\it Theorem}$\,$  A \s\   $(M,g)$ cannot be
timelike and null geodesically complete if the following are satisfied:
\medskip
\noindent(1) $R_{ij}K^iK^j\ge0$ for all non-spacelike vectors $K^i$;

\noindent(2) the generic condition is satisfied, that is,
every non-spacelike geodesic
contains a point at which $K_{[i}R_{j]el[m}K_{n]}K^e K^l\ne0$, where $K$
is the tangent to the nonspacelike geodesic;   

\noindent(3) the chronology condition holds on \m; that is, there are
no closed timelike curves in the spacetime, and,

\noindent(4) there exists in \m\  either a compact achronal set
(i.e. a set no two points of which are timelike related)
without edge or a closed
trapped surface, or a point $p$ such that for all
past directed null geodesics
from $p$, eventually $\theta$ must be negative.
\medskip

The main idea of the proof is the following. One shows that
the following three cannot hold simultaneously:
\medskip
\noindent(a) every inextendible non-spacelike geodesic contains
pairs of conjugate points;

\noindent(b) the chronology condition holds on \m;

\noindent(c) there exists an achronal set $\cal S$ in \m\ 
such that $E^+({\cal S})$ or $E^-({\cal S})$ is compact.

In the above, $E^{+}$ and $E^{-}$ indicate the future and
past horismos for the set $\cal S$ (for further definitions and detail
we refer to
Hawking and Ellis (1973), or Joshi (1993).

We note that while the geodesic incompleteness, as a definition
of spacetime singularities, allows various theorems to be proved on
the existence of singularities, it does not capture all possible
singular behaviors for a spacetime. It also does not imply that the
singularity predicted is necessarily a physically relevant powerful
curvature singularity. It does of course include many cases where that
will be the case. We discuss below such a scenario and the criterion 
for the singularity to be physically relevant and important.

\section{Strong Curvature Singularities}    
As we see above, the existence of an incomplete nonspacelike
geodesic or the existence of an inextendible nonspacelike
curve which has a finite length as measured by a generalized
affine parameter, implies the existence of a
spacetime singularity.
The {\it generalized affine length} for such a curve
is defined as (Hawking and Ellis, 1973),
$$ L(\lambda)=\int_0^a \left[\sum_{i=0}^3 (X^i)^2\right]^{1/2} ds$$
which is a finite quantity. The  $X^i$s are the components of the tangent
to the curve in a parallel propagated tetrad frame along the curve.
Each such incomplete curve defines a boundary point of the \s\  which is a \p.

The important point now is, in order to call this a genuine physical
singularity, one would typically like to associate  such a singularity with
unboundedly growing \s\  curvatures.
If all the curvature components and the scalar polynomials formed out of the
metric and the Riemann curvature tensor remained finite and well-behaved
in the limit of approach to the \p\ along an incomplete non-spacelike curve,
it may be possible to remove such a \p\ by extending the \s\  when the
differentiability requirements are lowered
(Clarke, 1986).

There are several ways in which such a requirement can be formalized.
For example, a {\it parallely propagated curvature singularity} is the one
which is the end point of at least one nonspacelike curve on which the
components of the Riemann curvature tensor are unbounded in a parallely
propagated frame. On the other hand, a {\it scalar polynomial \p} is the one
for which a scalar polynomial in the metric and the Riemann tensor takes an
unboundedly large value along at least one nonspacelike curve which has the
singular end point. This includes the cases such as the Schwarzschild
singularity where the Kretschmann scalar $R^{ijkl}R_{ijkl}$ blows up in the
limit as $r\to 0$.

What is the guarantee that such curvature singularities will at all occur
in general relativity? The answer to this question for the case of
parallely propagated curvature singularities is provided by a theorem of
Clarke (1975)
which establishes that for a globally hyperbolic \s\  \m\  which is
$C^{0-}$ inextendible, when the Riemann tensor is not very specialized in the
sense of not being type-D and electrovac at the singular end point, then the
singularity must be a parallely propagated curvature singularity.

Curvature singularities to be characterized below, also
arise for a wide range of spacetimes involving gravitational collapse.
This physically relevant class of singularities, called the 
{\it strong curvature singularities} was defined and analyzed by
Tipler (1977); Tipler, Clarke and Ellis (1980),
and
Clarke and Kr\'olak (1986).
The idea here is to define a physically all embracing
strong curvature \p\ in such a way so that all the objects falling within the
\p\ are destroyed and crushed to zero volume by the infinite
gravitational tidal forces.
The extension of spacetime becomes meaningless for such a strong singularity   
which destroys to zero size all the objects terminating at the singularity.
From this point of view, the strength of singularity may be considered
crucial to the issue of classically extending the spacetime, thus avoiding
the singularity. This is because for a strong curvature singularity
defined in the above sense, no continuous extension of the spacetime
may be possible.

\section{Can we avoid spacetime singularities?}
Given the scenario above, it is now clear that
spacetime singularities are an inevitable feature for most
of the physically reasonable models of universe and gravitational
systems within the framework of the Einstein theory of gravity. 
It is also seen that near such a
spacetime singularity, the classical description that predicted
it must itself breakdown. The existence of singularities in most of
the classical theories of gravity, under reasonable physical
conditions, imply that in a sense the Einstein gravity itself predicts
its own limitations, namely that it predicts regions of universe where
it must breakdown and a new and revised physical description must take
over.

As the curvatures and all other physical quantities 
must diverge near such a singularity, the quantum effects associated with
gravity are very likely to dominate such a regime.
It is possible that these may resolve the classical 
singularity itself. But we have no viable and consistent quantum
theory of gravity
available as of today despite serious attempts. Therefore
the issue of resolution of singularities as
produced by classical gravity remains open.

The other possibility is of course that
some of the assumptions of the singularity theorems
may be violated so as to avoid the singularity occurrence. 
Even when these are fairly
general, one could inquire whether some of these could actually breakdown and
do not hold in physically realistic models. This could save
us from the occurrence of singularities at the classical level
itself. Such possibilities
mean a possible violation of causality in the spacetime,
or no trapped surfaces occurring in the dynamical evolution
of universe, or possible violation of energy conditions.

The singularity theorem stated above and also
other singularity theorems contain the assumption of
causality or strong causality, or some other suitable
causality condition. Then the
alternative is that causality may be violated rather than a \p\
occurring in the \s .
So the implication of the \p\ theorem stated above is
when there is enough matter present in the universe, either the causality
is violated or a boundary point must exist for the \s.
In the cosmological case, such stress-energy density
will be provided by the microwave background
radiation, or in the case of stellar collapse
trapped surfaces may form
(Schoen and Yau, 1983), 
providing a condition
leading to the formation of a singularity.

The Einstein equations by themselves
do not rule out causality violating configurations which really depend on the
global topology of the \s. Hence the question of causality violations versus
\s\  \p\ needs a careful examination as to whether causality violation
could offer an alternative to singularity formation. Similarly, it
also has to be inquired if the violation of energy conditions or
non-occurrence of trapped surfaces may be realized so as to achieve
the singularity avoidance in a spacetime. We discuss briefly some of
these points below.

\section{Causality violations}
The causal structure in a spacetime specifies what events
can be related to each other by means of timelike or light signals.
A typical causality violation would mean that an event 
could be in its own past, which is contrary to our normal
understanding of time, and that of past and future.
This has been examined
in considerable detail in general relativity, and that 
there is no causality violation taking place in the spacetime 
is one of the important assumptions used by singularity 
theorems. However,  
general relativity allows for situations where causality violation
is permitted in a \s . The G\"odel solution 
(G\"odel, 1949)
allows the existence of a closed timelike curve through 
every point of the \s .

One would of course like to rule out if possible
the causality violations on physical grounds, treating them 
as a very pathological behaviour in that in such a case one
would be able to enter one's own past. However, as they are
allowed in principle in general relativity, so one must
rule them out only by an additional assumption.
The question then is, can one avoid the \s\  
singularities if one allows for the violation of causality? 
This has been considered by researchers 
and it is seen that the causality violation 
in its own right creates \s\  singularities again under 
certain reasonable conditions. Thus, this path of avoiding 
spacetime singularities does not appear to be very promising.

Specifically, the question of finite causality violations 
in asymptotically flat spacetime was examined by 
Tipler (1976, 1977).
This showed that the causality violation   
in the form of closed timelike lines is necessarily accompanied 
by incomplete null geodesics, provided the strong energy 
condition is satisfied for all null vectors and if 
the generic condition is satisfied. It was assumed
that the energy density $\rho$ has a positive minimum 
along past directed null geodesics.

There is in fact a heirarchy of causality conditions 
available for a spacetime. It may be causal in the sense of
having no closed nonspacelike curves. But given an event,
future directed nonspacelike curves from the same 
could return to its arbitrarily close neighborhood
in the spacetime. This is as bad as a causality violation
itself. The higher order causality conditions such as
strong causality and stable causality rule out such 
behaviour. 
Of the higher order causality conditions, much physical
importance is attached to the stable causality which 
ensures that if \m\  is causal, its causality should not 
be disturbed with small perturbations in the metric tensor. 
Presumably, the general relativity is a classical approximation 
to some, as yet unknown, quantum theory of gravity in which 
the value of the metric at a point will not be exactly known 
and small fluctuations in the value must be taken into account.

Results on causality violations and higher-order 
causality violations with reference to occurrence of singularities 
were obtained by 
Joshi (1981), and 
Joshi and Saraykar (1986), 
who show that the causality violations must be accompanied by  
singularities even when the \s\  is causal but the higher order 
causality conditions are violated.
Thus we know that for a causal \s\,  the violations of 
higher order causality conditions give rise to \s\ singularities. 
Another question examined was that of measure of causality
violating sets when such a violation occurs. It turns out that 
in many cases, the causality violating sets in a \s\  will 
have a zero measure, and thus such a causality violation 
may not be taken very seriously. Also, 
Clarke and Joshi (1988) 
studied global causality violation for a reflecting
spacetime and the theorems of 
Kriele (1990) 
improved some of the conditions under which the  
results on chronology violations implying the singularities 
have been obtained. Also, global causality violating \s s 
were studied by 
Clarke and de Felice (1982).
What we discussed above implies that 
if the causality of \m\  breaks down with the slightest
perturbation of the metric then this must be accompanied 
by the occurrence of \s\  singularities.

As a whole, the above results imply that violating either 
causality or 
any of the higher-order causality conditions may not be 
considered a good alternative to the occurrence of  \s\  
singularities. There are also philosophical
problems connected with the issue of causality violation 
such as entering one's own past. But even if one allowed
for the causality violations, the above results show 
that these are necessarily accompanied by spacetime 
singularities again.

\section{Energy conditions and trapped surfaces}
Another possibility to avoid singularities is a possible
violation of energy conditions. This is another of the assumptions 
in the proofs for singularity theorems. In fact, this possibility 
has also been explored in some detail and it turns out that 
as long as there is no gross or very powerful violation of 
energy conditions over global regions in the universe, this 
would not help avoid singularities either.

For example, the energy condition could be violated 
locally at certain spacetime points, or in certain regions 
of spacetimes due to peculiar physics there. But as long as it 
holds on an average, in the sense that the stress-energy density 
is positive in an integrated sense, then still spacetime
singularities do occur (for a discussion and references, see e.g.
Joshi, 1993).

On a global scale, there is evidence now that the universe 
may be dominated by a dark energy field. There is no clarity 
as to what exactly such a field would be and what would be its 
origin. It could be due to scalar fields or ghost fields floating
in the universe, or due to a non-zero positive cosmological
constant present in the Einstein equations. In such a case, 
the weak or strong energy conditions may be violated 
depending on the nature of these exotic fields. However, 
in the earlier universe of a matter dominated phase, the 
positive matter fields would again dominate, thus respecting 
the energy conditions even if they are violated at the 
present epoch.

Again, the above discussion is in the context of a 
cosmological scenario. When it comes to the gravitational 
collapse of massive stars, clearly their density and overall
energy content are dominated by the ordinary matter fields 
we are much more familiar with. Such matter certainly respects
the energy conditions modulo some minor violations if at all
any. Thus for gravitational collapse of massive stars, one
would expect the energy conditions to hold and the conclusions
on singularity occurrence stated above would apply.

Yet another possibility to avoid singularity is to avoid 
trapped surfaces occurring in the spacetime. Indeed, such a 
route can give rise to geodesically complete spacetimes, 
as was shown by 
Senovilla (1990).
As for the cosmological scenario, basically this means
and amounts to the condition that the matter energy densities
must fall off sufficiently rapidly on any given spacelike 
surface, and in an averaged sense, in order to avoid the 
cosmological trapped surfaces.
Whether such a condition is realizable in the universe 
would have to be checked through observational tests.
A sufficiently uniform energy density, such as say the 
microwave background radiation could in turn cause
the cosmic trapping. As for the massive stars, the densities
are of course very high indeed and would only grow 
for example in a gravitational collapse. Therefore in 
collapse scenarios, the trapped surfaces are unlikely
to be avoided.

Further to the considerations such as above, if we accept 
on the whole that spacetime singularities do 
occur under fairly general conditions in the framework of the Einstein
theory of gravity, or for classical gravity in general, then
one must consider physical implications and consequences of such a 
scenario for physics and cosmology. As we noted earlier,
two main arenas of physical relevance where spacetime singularities
will be of interest are the cosmological situation and the 
gravitational collapse scenarios.

\section{Fundamental implications and challenges}
The existence of spacetime singularities in Einstein 
gravity and other classical theories of gravitation poses intriguing 
challenges and fundamental questions 
in physics as well as cosmology. These would have farreaching 
consequences on our 
current understanding of the universe and how we try to 
further model it, as we shall try to bring out in 
rest of this article.

The inevitable existence of singularities for 
wide classes of rather general models of spacetimes means that
the classical gravity evolutions necessarily give rise to 
regions in the spacetime universe where the densities and 
spacetime curvatures would really grow arbitrarily high 
without any bounds, and where all other relevant physical 
parameters also would diverge.

To take the first physical scenario, such a phenomenon 
in cosmology would correspond to a singularity that will represent
the origin of the universe.
Secondly, whenever locally a large quantity of matter and energy
collapses under the force of its own gravity, a singularity will
occur. This later situation
will be effectively realized in the gravitational collapse
of massive stars in the universe, which collapse and shrink 
catastrophically under their self-gravity, when the star has
exhausted its nuclear fuel within that earlier supplied the 
internal pressures to halt the infall due to gravity.

Over past decades, once the existence of spacetime singularities
was accepted, there have been major efforts to understand the physics 
in the vicinity of the same. In the cosmological case, this has 
given rise to an entire physics of the early universe, which depicts 
the few initial moments immediately after the big bang singularity 
from which the universe is supposed to have emerged. The 
complexities in this case have been enormous, both physicswise, 
and conceptually. The physics complexities arise because when
trying to understand physics close to the hot big bang singularity,
we are dealing with the highest energy scales, never seen earlier
in any laboratory physics experiments. Our particle physics
theories are then to be stretched to the extreme where there is 
no definite or unique framework available to deal with
these phenomena. Understanding early universe physics has of course
very big consequences in that it governs the most important 
physical phenomena 
such as the later galaxy formation in the universe, and other 
issues related to the large scale structure of the universe.

As for the conceptual issues, the simultaneous big bang singularity
gives rise to a host of problems and puzzles. One of these is the 
`horizon problem' which arises due to the causal structure of this
spacetime. Distinct regions of the universe simply cannot interact 
with each other due to the cosmic horizons and it becomes extremely
difficult to explain the average overall current homogeneity of
the universe on a large enough scale. There are also other issues 
such as why the current
universe is looking so flat, which is called the `flatness' problem.     
As a possible resolve to these dilemmas, the inflationary models
for early universe have been proposed, various facets of which 
are still very much under an active debate.

The key issue, as far as the big bang singularity is concerned, is
that it happened only once in the past and there is no way to
probe it any further other then current observations on universe
and their extrapolation in the past. One must observe deeper
and deeper into the space and back into time to understand the 
nature and physics of this early universe singularity.

As we mentioned above, the other class of such spacetime
singularities will occur in the gravitational collapse of massive
stars in the universe. Unlike the big bang, such a singularity 
will occur whenever a massive star in the universe collapses.
This is therefore more amenable to observational tests.

There are rather fundamental cosmic conundrums associated
with singularities of gravitational collapse. One of the most
intriguing ones of these is the question whether such a singularity 
will be visible to external faraway observers in the universe.
The big bang singularity is visible to us in principle, as we
get to see the light from the same. But as we discuss below, 
the singularity of collapse can sometimes be hidden below
the event horizons of gravity, and therefore not visible.
The possibility that all singularities of collapse will be
necessarily hidden inside horizons is called the {\it cosmic 
censorship conjecture}.
As we discuss below this is not yet proved, and in fact 
singularities of collapse can be visible under many physical
circumstances.

When visible or naked singularities develop in gravitational 
collapse, they give rise again to extremely intriguing physical 
possibilities and problems. The opportunity offered in that 
case is, we may have the possibility to observe the possible
ultra-high energy physical processes occurring in such a region
of universe, including the quantum gravity effects. Such 
observations of ultra-high energy events in the universe could
provide observational tests and guide our efforts for a possible
quantum theory of gravity. But a conundrum that is presented 
is, whether this would break the so called {\it `classical 
predictability'} of the spacetime universe. We shall discuss 
this further below. On the other hand, even when the singularity
is necessarily hidden within a black hole, that still gives
rise to profound puzzles such as the {\it `information paradox'},
issues with {\it unitarity} and such other problems. So the 
point is, even if the cosmic censorship was correct and all 
singularities were hidden inside black holes only, still we 
shall be faced with many deep paradoxes, which are not 
unique to naked singularities only.

It would be only reasonable to say that all these deep 
physical as well as conceptual issues are closely connected with
the existence and formation of spacetime singularities in the 
dynamical gravitational processes taking place in the universe.
While the big bang singularity happened only once in the 
past, the singularities of collapse have in fact a repeated 
occurrence, and hence they possess an interesting observational 
perspective and potential. We shall therefore discuss the 
same in some detail below, while also providing the key 
ingredients of black hole physics in the process.

\section{Gravitational collapse}
When a massive star more than about ten solar masses 
exhausted its internal nuclear fuel, it is believed to enter
the stage of a continual gravitational collapse without 
any final equilibrium state. The star then goes on shrinking 
in its radius, reaching higher and higher densities. What 
would be the final fate of such an object according to 
the general theory of relativity? This is one of the central 
questions in relativistic astrophysics and gravitation theory 
today. It is suggested that the ultra-dense object that 
forms as a result of the collapse could be a black hole in 
space and time from which not even light rays escape. 
Alternatively, if an event horizon of gravity fails to 
cover the final super ultra-dense crunch, it could be a visible 
singularity in the spacetime which could causally interact 
with the outside universe and from which region the emissions 
of light and matter may be possible.

The issue is of importance from both the theoretical 
as well as observational point of view. At the theoretical level, 
working out the final fate of collapse in general relativity 
is crucial to the problem of asymptotic predictability, 
namely, whether the singularities forming as the collapse endstate 
will be necessarily covered by the event horizons of gravity. 
Such a {\it censorship hypothesis} remains fundamental to the 
theoretical foundations of black hole physics and its many 
recent astrophysical applications. These include the area theorem 
for black holes, laws of black hole thermodynamics, Hawking 
radiation, predictability in a spacetime, and on observational 
side the accretion of matter by black holes, massive black 
holes at the center of galaxies etc. On the other hand, existence 
of visible or naked singularities offer a new approach on 
these issues requiring modification and reformulation 
of our usual theoretical conception on black holes.

We mention and discuss below some of these recent developments
in these directions, examining the possible final fate of gravitational
collapse. To investigate this issue, dynamical collapse scenarios 
have been examined in the past decade or so for many cases such 
as clouds composed of dust, radiation, perfect fluids,
or matter with more general equations of state 
(see e.g. Joshi 2008 
for references and details).
We discuss these developments and the implications for 
a possible formulation of cosmic censorship are indicated, 
mentioning the open problems in the field.

\section{Spherical collapse and the black hole}
To understand the final fate of a massive gravitationally
collapsing object we first outline here
the spherically symmetric collapse situation.
Though this is an idealization, the advantage is one can
solve the case analytically to get exact
results when matter is homogeneous dust.
In fact, the basic motivations for the idea and theory 
of black holes come from this collapse model,
first worked out by 
Oppenheimer and Snyder (1939)
and Datt (1938).

Consider a gravitationally collapsing spherical 
massive star. The interior solution for the object 
depends on the properties of matter, equation of state, 
and the physical processes taking place within the stellar 
inside. But assuming the matter to be
pressureless dust allows the problem to be solved analytically, 
providing important insights. The energy-momentum tensor
is given by $T^{ij}= \rho u^iu^j$, and the Einstein
equations are to be solved for the spherically symmetric 
metric. The metric potentials can be solved and the 
interior geometry of the collapsing dust ball 
is given by,
$$ ds^2= -dt^2+ R^2(t)\left[ {dr^2\over 1-r^2}+ r^2 d\Omega^2
  \right]$$
where $d\Omega^2=d\theta^2+ sin^2\theta d\phi^2$ is the 
metric on two-sphere. The interior geometry of cloud matches 
at the boundary $r=r_b$ with the exterior Schwarzschild 
spacetime.

\begin{center}
\leavevmode\epsfysize=3.5 in\epsfbox{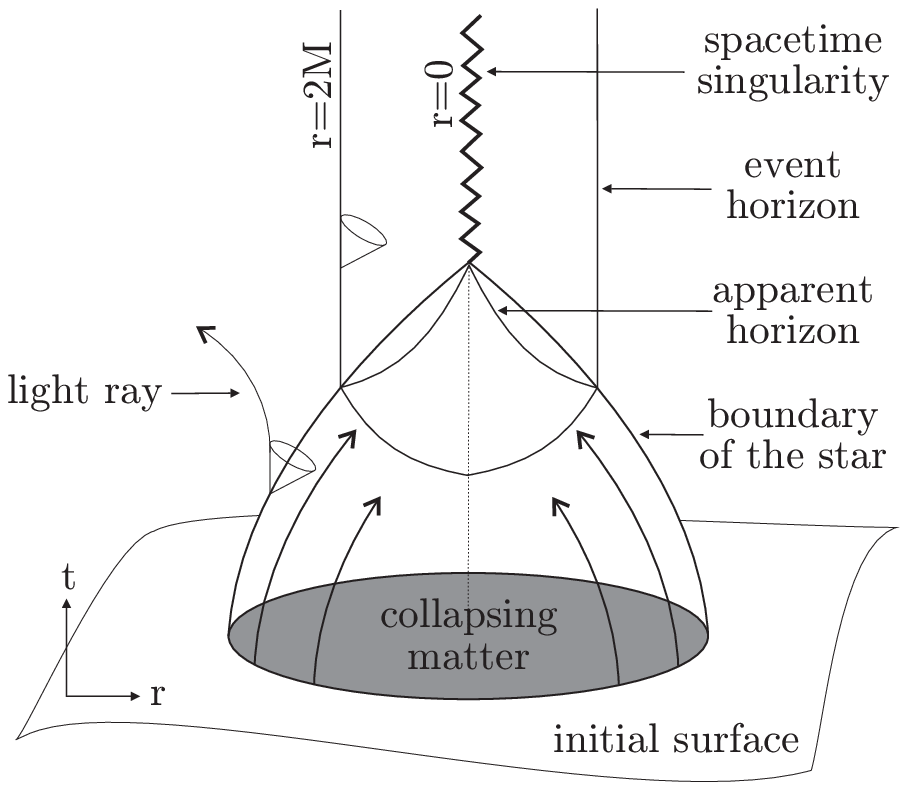}
\end{center}
\noindent {\small Fig 1: The gravitational collapse of a 
spherically symmetric homogeneous dust cloud. The event horizon
forms prior to the singularity and the collapse endstate 
is a black hole.}

The basic features of such a configuration are given 
in Fig 1.  The collapse is initiated when the star surface 
is outside the Schwarzschild radius $r = 2m$ and a light 
ray from the surface of the star can escape to infinity.
But once the star has collapsed below $r = 2m$, a black 
hole region of no escape develops in the spacetime which
is bound by the event horizon at $r =2m$.  Any point in 
this empty region represents a {\it trapped surface}, 
a two-sphere for which both the outgoing and ingoing families 
of null geodesics emitted from this point converge and 
so no light comes out of this region. Then, the collapse 
to an infinite density and curvature singularity at 
$r =0$ becomes inevitable in a finite proper time as 
measured by an observer on the surface of the star. The 
black hole region in the resulting vacuum Schwarzschild geometry 
is given by $0 < r < 2m$ with the event horizon as the outer 
boundary at which the radial outwards photons stay where 
they are but all the rest are dragged inwards. No
information from the black hole can propagate outside $r=2m$ 
to observers far away. We thus see that the collapse gives 
rise to a black hole in the spacetime which covers the
resulting spacetime singularity. The ultimate fate of 
the star undergoing such a collapse is an infinite 
curvature singularity at $r = 0$, completely hidden within 
the black hole. No emissions or light rays from the singularity 
go out to observer at infinity and the singularity is 
causally disconnected from the outside spacetime.

The question now is whether one could generalize these 
conclusions on the occurrence of spacetime singularity in 
collapse for more general forms of matter or for non-spherical 
situations, or possibly for small perturbations away from spherical 
symmetry.  It is known using the stability of Cauchy development 
in general relativity that the formation of trapped surfaces 
is indeed a stable property when departures from spherical symmetry 
are taken into account.  The argument essentially is the 
following: Considering a spherical collapse evolution from given 
initial data on a partial Cauchy surface $S$, we find the 
formation of trapped surfaces $T$ in the form of all the
spheres with $r < 2m$ in the exterior Schwarzschild geometry.  
The stability of Cauchy development then implies that for 
all initial data sufficiently near to the original data 
in the compact region $J^{+} (S) \cap J^{-} (T)$, where
$J^{+}$ and $J^{-}$ denote the causal futures or pasts of $S$
respectively, the trapped surfaces still must occur.  Then, 
the curvature singularity of spherical collapse also turns 
out to be a stable feature, as implied by the singularity theorems
which show that the closed trapped surfaces always imply the 
existence of a spacetime singularity under reasonable 
general enough conditions.

\section{Cosmic censorship hypothesis}
The real stars in the universe are not made of pressureless 
homogeneous dust. They are inhomogeneous, typically with density 
higher at center,
may have non-trivial matter with an equation of state as
yet unknown, and there is spin. Will a physically realistic
collapse of such a star necessarily end up in the black hole final
state only, just as in the idealized case of the Oppenheimer-
Snyder-Datt model above? In other words, while the more general
gravitational collapse will also end up in a spacetime singularity,
the question is whether the singularity will be again necessarily 
covered inside an event horizon of gravity.

In fact, there is no proof available that such a singularity 
will continue to be hidden within a black hole and remain causally 
disconnected from outside observers, even when the collapse 
is not exactly spherical or when the matter does not have the 
form of exact homogeneous dust. Therefore, in order to generalize 
the notion of black holes to more general gravitational collapse 
situations, it becomes necessary to rule out such naked or visible
singularities by means of an explicit assumption.  This is 
stated as the {\it cosmic censorship hypothesis}, which essentially 
says that if $S$ is a partial Cauchy surface from which the 
collapse commences, then there are no naked singularities to the 
future of $S$ which could be seen from the future null infinity. 
This is true for the spherical homogeneous dust collapse, where 
the resulting spacetime is future asymptotically predictable and 
the censorship holds. In such a case, the breakdown of physical 
theory at the spacetime singularity does not disturb prediction
in future for the outside asymptotically flat region.

The corresponding scenario for other more general 
collapse situations, when inhomogeneities or non-sphericity and
such other physically realistic features are allowed for
has to be investigated. 
The answer in general is not known either as a proof of the future
asymptotic predictability for general spacetimes or in the form
of any general theorem on cosmic censorship.
It is clear that the assumption of censorship in a suitable 
form is crucial to the basic results in black hole physics. Actually
when one considers the gravitational collapse in a generic situation, 
the very existence of black holes requires this hypothesis.

To establish the censorship by means of a rigorous proof
certainly requires a much more precise formulation of the hypothesis.
The statement that result of complete gravitational collapse must
be a black hole only and not a naked singularity, or all singularities 
of collapse must be hidden inside black holes, is not rigorous
enough. Because under general circumstances, the censorship or 
asymptotic predictability is false as one could always choose a 
spacetime manifold with a naked singularity which would be a solution 
to Einstein's equations if we define $T_{ij} \equiv (1/8\pi)G_{ij}$.
So certain conditions on the stress-energy tensor
are required at the minimum, say for example, an energy condition.  
However, to obtain
an exact set of conditions on matter fields to prove the 
censorship hypothesis turns out to be an extremely difficult 
task as yet not accomplished.

The requirements in the black hole physics and general
predictability have led to several different possible formulations 
of cosmic censorship hypothesis, none of which proved as yet.  
The {\it weak cosmic censorship}, or asymptotic predictability, 
postulates that the singularities of gravitational collapse 
cannot influence events near the future null infinity. 
The other version called the {\it strong cosmic censorship}, 
is a general predictability requirement on any spacetime, 
stating that all physically reasonable spacetimes must be 
globally hyperbolic 
(see e.g. Penrose, 1979).
The {\it global hyperbolicity} here means that we must
be able to predict the entire future and past evolutions in 
the universe by means of the Einstein equations, given the initial 
data on a three-dimensional spacelike hypersurface in 
the spacetime.

On a further analysis, it however becomes clear that 
such formulations need much more sharpening if at all any 
concrete proof is to be obtained. In fact, as for the cosmic 
censorship, 
it is a major problem in itself to find a satisfactory and 
mathematically rigorous formulation of what is physically
desired to be achieved. Presently, there is no general proof
available for any suitably formulated version of the weak 
censorship. The main difficulty seems to be that the event horizon 
is a feature depending on the whole future behavior of the 
solution over an infinite time period, but the present theory 
of quasi-linear hyperbolic equations guarantee the existence 
and regularity of solutions over only a finite time internal.  
It is clear that even if true, any proof for a suitable version 
of weak censorship seems to require much more knowledge
on general global properties of Einstein equations 
than is known currently.

To summarize the situation, cosmic censorship 
is clearly a crucial assumption underlying all of the 
black hole physics and gravitational collapse theory and 
related important areas. The first major task here would be 
to formulate rigorously a satisfactory statement for cosmic 
censorship, which if not true would throw the black hole dynamics 
into serious doubt. That is why censorship is one of the most 
important open problems for gravitation theory today.  
No proof, however, seems possible unless some major 
theoretical advances by way of mathematical techniques and 
understanding global structure of Einstein equations are 
made, and direction for such theoretical advances
needed is far from clear at present.

We therefore conclude that the first
and foremost task at the moment is to carry out a 
detailed and careful examination of various gravitational 
collapse scenarios to examine for their end states.
It is only such an investigation of more general collapse
situations which could indicate what theoretical advances 
to expect for any proof, and what features to avoid  
while formulating the cosmic censorship.
Basically, we still do not have sufficient data and information
available on the various possibilities for gravitationally
collapsing configurations so as to decide one way or other on the
issue of censorship.

In recent years, many investigations have been carried 
out from such a perspective on gravitational collapse, either for 
inhomogeneous dust collapse or with more general matter fields. 
It turns out that the collapse outcome is not a black hole
always and the naked singularity final state can arise in
a variety of situations. In the next sections we discuss 
some of these developments.

\section{Inhomogeneous Dust Collapse}
Since we are interested in collapse,
we require that the spacetime contains a regular initial spacelike
hypersurface on which the matter fields as represented by the stress-energy
tensor $T_{ij}$, have a compact support and all physical quantities are
well-behaved on this surface. Also, the matter should
satisfy a suitable energy condition and the Einstein equations are
satisfied. We say that the spacetime contains a naked singularity
if there is a future directed nonspacelike curve which reaches a far
away observer at infinity in future, which in the past terminates
at the singularity.

As an immediate generalization of the Oppenheimer-Snyder-Datt
homogeneous dust collapse, one could consider the collapse of
inhomogeneous dust and examine the nature and structure of resulting
singularity with special reference to censorship and the occurrence
of black holes and naked singularities. The main motivation to
discuss this situation is this provides a clear picture in
an explicit manner of what is possible in
gravitational collapse. One could ask how the conclusions
given above for homogeneous collapse are modified when the
inhomogeneities of matter distribution are taken into account.
Clearly, it is important to include effects of inhomogeneities because
typically a physically realistic collapse would start from an inhomogeneous
initial data with a centrally peaked density profile.

This question of inhomogeneous dust collapse has attracted 
attention of many researchers and it is seen that the introduction of 
inhomogeneities leads to a rather different picture of gravitational collapse. 
It turns out that while homogeneous collapse leads to a black hole
formation, introduction of any physically realistic inhomogeneity,
e.g. the density peaked at the center of the cloud and slowly
decreasing away, leads to a naked singularity final state for
the collapse. This is certainly an intriguing result implying that
the black hole formation in gravitational collapse need not be 
such a stable phenomenon as it was thought to be the case.

The problem was investigated in detail using the
Tolman-Bondi-Lemaitre models, which describe gravitational collapse
of an inhomogeneous spherically symmetric dust cloud
(Joshi and Dwivedi 1993). 
This is an infinite dimensional family of asymptotically flat
solutions of Einstein equations, which is matched to the Schwarzschild
spacetime outside the boundary of the collapsing star. The
Oppenheimer-Snyder-Datt model is a special case of this class
of solutions.

It is seen that the introduction of inhomogeneities
leads to a rather different picture of gravitational collapse. 
The metric for spherically
symmetric collapse of inhomogeneous dust, in comoving
coordinates $(t, r, \theta,\phi)$, is given by,
$$ ds^2= -dt^2+{R'^2\over1+f}dr^2+R^2(d\theta^2+sin^2\theta\, d\phi^2)$$
$$T^{ij}=\epsilon \delta^i_t \delta^j_t,\quad \epsilon=\epsilon(t,r)={F'
\over R^2R'},$$
where $T^{ij}$ is the stress-energy tensor, $\epsilon$ is
the energy density, and $R$ is a function of both $t$ and $r$ given by
$$ \dot R^2={F\over R}+f $$
Here the dot and prime denote partial derivatives with respect
to the parameters $t$ and $r$ respectively. As we are considering collapse,
we require $\dot R(t,r)<0.$ The quantities $F$ and $f$ are arbitrary
functions of $r$ and $4\pi R^2(t,r)$ is the proper area of the  mass shells.
The physical area of such a shell at $r=\hbox{const.}$ goes 
to zero when $R(t,r)=0$.
For gravitational collapse situation,
we take $\epsilon$ to have compact support
on an initial spacelike hypersurface and the spacetime is matched
at some $r=\hbox{const.}=r_c$ to the exterior Schwarzschild field
with total Schwarzschild mass $m(r_c)=M$ enclosed within the dust ball
of coordinate radius of $r=r_c$. The apparent horizon in the interior
dust ball lies at $R=F(r)$.

Using this framework, the nature of the singularity $R=0$ can
be examined. In particular, the problem of nakedness or otherwise of
the singularity can be reduced to the existence of real, positive roots
of an algebraic equation $V(X)=0$, constructed out of the free 
functions $F$ and $f$ and their derivatives,
which constitute the initial data of
this problem. If the equation $V(X)=0$ has a real positive root, the singularity
could be naked. In order to be the end point of null geodesics at least
one real positive value of $X_0$ should satisfy the above.
If no real positive root of the above is found, the singularity
is not naked. It should be noted that
many real positive roots of the above equation may exist which give the
possible values of tangents to the singular null geodesics
terminating at the singularity in the past. Suppose now $X=X_0$ is
a simple root to $V(X)=0$.
To determine whether $X_0$ is realized as a tangent along any outgoing
singular geodesics to give a naked singularity, one can integrate the
equation of the radial null geodesics in the form $r=r(X)$
and it is seen that there is always at least one null geodesic terminating
at the singularity $t=0,r=0$, with $X=X_0$. In addition there would be
infinitely many integral curves as well, depending on the values of the
parameters involved, that terminate at the singularity.
It is thus seen that the existence of a positive real root of
the equation $V(X)=0$ is a
necessary and sufficient condition for the singularity to be
naked. Finally, to determine the curvature strength of the naked singularity
at $t=0$, $r=0$, one may analyze the quantity
$ k^2 R_{ab}K^aK^b$ near the singularity. Standard analysis shows that
the strong curvature condition is satisfied, in that the above
quantity remains finite in the limit of approach to the singularity.
The spacetime picture for a collapse terminating in a naked 
singularity is given in Fig 2.

\begin{center}
\leavevmode\epsfysize=3.5 in\epsfbox{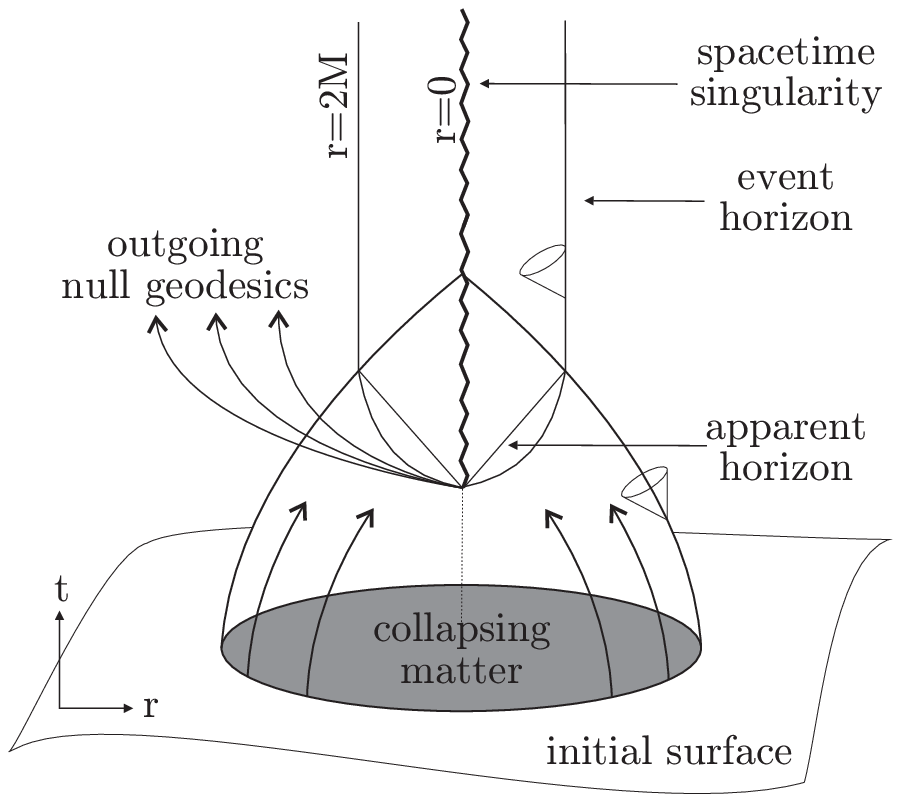}
\end{center}
\noindent {\small Fig 2: The gravitational collapse of a 
spherical but inhomogeneous dust cloud with a density profile peaked 
at the center. The event horizon no longer forms prior to the 
singularity and the collapse endstate is a naked singularity.}

The assumption of vanishing pressures,
which could be important in the final stages of the collapse, may be
considered as the limitation of dust models.
On the other hand, it is also argued sometimes that in the final stages
of collapse the dust equation of state could be relevant
and at higher and higher densities the matter may behave
much more like dust. Further, if there are no large negative pressures
(as implied by the validity of the energy conditions), then the pressure
also might contribute gravitationally in a positive manner
to the overall effect of dust and may not alter the 
final conclusions.

\section{Collapse with general matter fields}
It is clearly important to consider collapse situations which consider
matter with non-zero pressures and with reasonable equations of state.
It is possible that pressures may play an important role for the later
stages of collapse and one must investigate the possibility if pressure
gradients could prevent the occurrence of naked singularity.

Many collapse scenarios have been considered by now with
non-zero pressures and physically reasonable equations of state.
What one needs to examine here again is the existence,
the termination of future directed nonspacelike geodesic families
at the singularity in the past, and the strength of such a singularity 
for collapse with non-zero pressure.

A useful insight into this issue is provided by self-similar 
collapse for a perfect fluid with a linear equation of state
$p=k\rho$. A numerical treatment
of self-similar perfect fluid collapse was given by 
Ori and Piran (1987) 
and the analytic consideration for the same was given by
Joshi and Dwivedi (1992).
It is seen that the collapse evolutions as allowed 
by the Einstein equations permit for both the black hole and
naked singularity final states. 
If in a self-similar collapse a single null radial geodesic 
escapes the singularity, then in fact an entire family of nonspacelike 
geodesics would also escape provided the positivity of energy 
density is satisfied. It also follows that no families of nonspacelike
geodesics would escape the singularity, even though a single null
trajectory might, if the weak energy condition is violated.
The singularity will be globally visible to faraway observers
in the spacetime for a wide set of conditions. 
What these results show is, naked singularity is not avoided
by introduction of non-zero pressures or a reasonable equation
of state.

Actually, consideration of matter forms such as directed 
radiation, dust, perfect fluids etc imply a similar general pattern 
emerging as far as the final outcome of collapse is concerned. 
Basically, the result that emerges is, depending on the nature
of the regular initial data in terms of the density and pressure
profiles, the Einstein equations permit both the classes 
of dynamical evolutions, those leading to either of the black
hole or naked singularity final states.

Hence one could ask the question
whether the final fate of collapse would be independent of the form of
the matter under consideration. An answer to this is useful
because it was often thought 
that once a suitable form of matter with an appropriate 
equation of state, also satisfying energy conditions is considered 
then there may be no naked singularities. 
Of course, there is always a possibility that
during the final stages of collapse the matter may not have 
any of the forms such as dust or perfect fluids considered above, 
because such relativistic 
fluids are phenomenological and perhaps one must treat matter 
in terms of some fundamental field, such as for example, 
a massless scalar field. In that context, a naked singularity
is also seen to form for the scalar field collapse
(Choptuik, 1993), 
though for a fine-tuned initial data.

In the above context, efforts in the direction of understanding
collapse final states for general matter fields are worth mentioning, 
which generalize the above results on perfect fluid 
to matter forms without any restriction on the form of $T_{ij}$, 
with the matter satisfying the weak energy condition. 
A consideration to a general form of matter was given by 
Lake (1992); 
and by 
Szekeres and Iyer (1993),
who do not start by assuming an equation of state but
a class of metric coefficients is considered with a certain power law
behavior. Also, 
Joshi and Dwivedi (1999) and Goswami and Joshi (2007), 
gave results in this direction. The main argument 
is along the following lines. It 
was pointed out above that naked singularities could 
form in gravitational collapse from a regular initial data, 
from which non-zero measure families of nonspacelike trajectories
come out. The criterion for the existence of such singularities
was characterized in terms of the existence of real positive roots of
an algebraic equation constructed out of the field variables. 
A similar procedure is developed now for general form of matter. 
In comoving coordinates, the general matter can be described 
by three free functions, namely the energy density and the radial 
and tangential pressures. The existence of naked singularity is 
again characterized in terms of the real positive roots of 
an algebraic equation, constructed from the equations of nonspacelike 
geodesics which involve the three metric functions. The field 
equations then relate these metric functions to the matter 
variables and it is seen that for a subspace of this free 
initial data in terms of matter variables, the above algebraic 
equation will have real positive roots, producing a naked 
singularity in the spacetime. When there are no such roots
existing, the endstate is a black hole.

It follows that the occurrence or otherwise of the 
naked singularity is basically related to the choice of 
initial data to the Einstein field equations as determined
by the allowed evolutions. Therefore these occur from  
regular initial data within the general context considered, subject 
to the matter satisfying weak energy condition. The condition 
on initial data which leads to the formation of black hole 
is also characterized.

It would then appear that the occurrence of naked 
singularity or a black hole is more a problem of choice of 
the initial data for field equations rather than that 
of the form of matter or the equation of state (see Fig 3). 

\begin{center}
\leavevmode\epsfysize=3.5 in\epsfbox{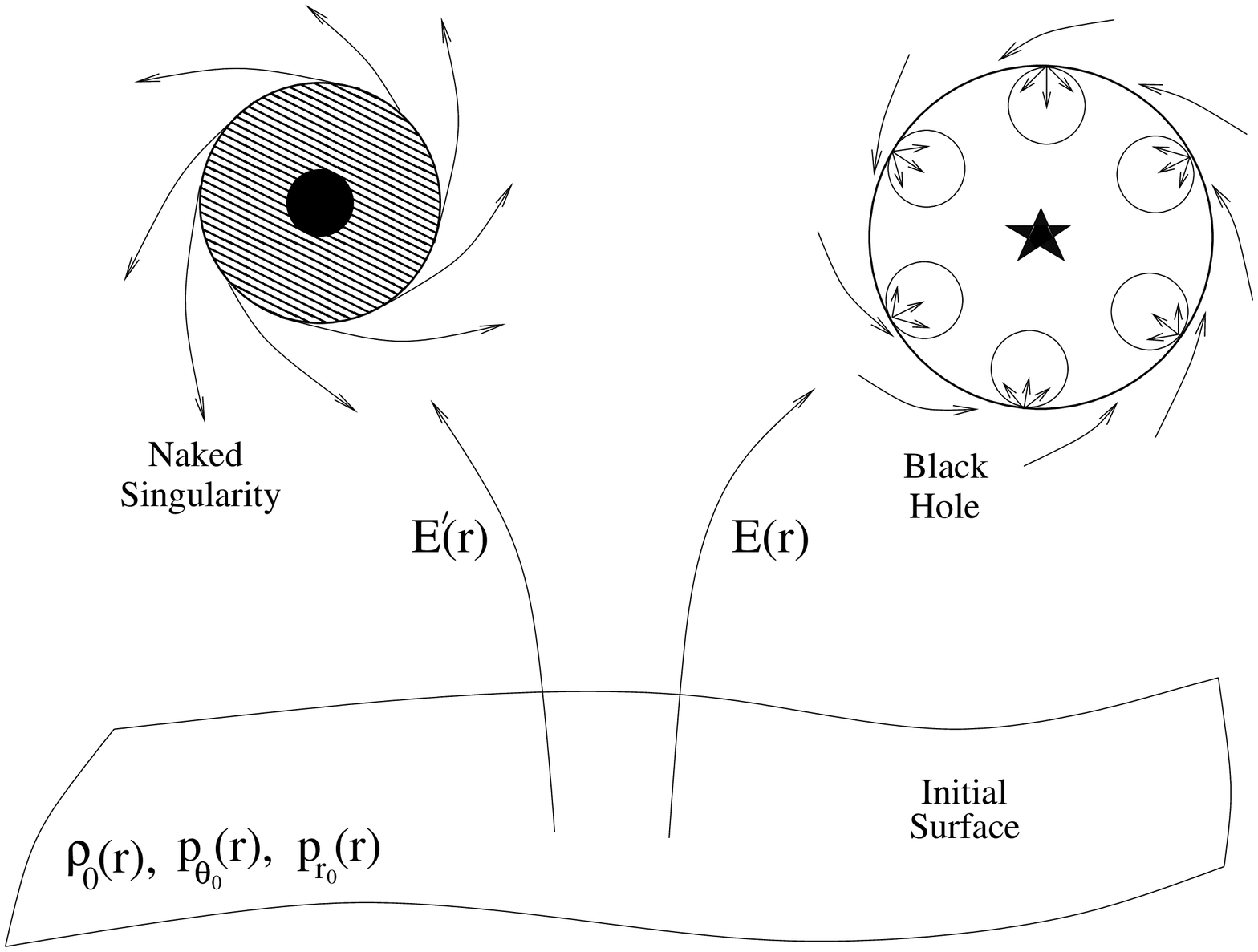}
\end{center}
\noindent {\small Fig 3: For a generic collapse of a 
general matter field, the collapse final state can be either 
a black hole or a naked singularity depending on the dynamical 
evolution chosen as allowed by the Einstein equations.}

Such a conclusion has an important implication for cosmic 
censorship in that in order to preserve the same one has to avoid 
all such regular initial data causing naked singularity, and hence 
a much deeper understanding of the initial data space is required 
in order to determine such initial data and the kind of physical 
parameters they would specify. In other words, this classifies 
the range of physical parameters to be avoided for a particular
form of matter. Such an understanding would also pave the way 
for black hole physics to use only those ranges of allowed 
parameter values which produce black holes only, thus putting 
black hole physics on a more firm footing.

\section{Non-spherical Collapse and numerical simulations}
Basically, the results and detailed studies such as above 
on gravitational collapse show that the cosmic censorship 
cannot hold in an unqualified general form. It must be properly 
fine-tuned and only under certain suitably restrictive conditions 
on collapse the black holes will form.

An important question at the same time is: What will be 
the final fate of gravitational collapse which is not
spherically symmetric? The main phases of spherical collapse of a
massive star would be typically instability,
implosion of matter, and subsequent formation of an event horizon and
a spacetime singularity of infinite density and curvature with infinite
gravitational tidal forces. This singularity may or may not be fully
covered by the horizon as we discussed above.

As noted, small perturbations over the sphericity 
would leave the situation unchanged in the sense that an 
event horizon will continue to form in the advanced stages 
of the collapse.
The next question then is, do horizons still form when the 
fluctuations from the spherical symmetry are high and the collapse 
is highly non-spherical? It was shown by 
Thorne (1972), 
for example, that when there is no spherical
symmetry, the collapse of infinite cylinders do give rise to naked
singularities in general relativity, which are not covered 
by horizons. This situation motivated Thorne to propose 
the {\it hoop conjecture} for finite systems in an asymptotically 
flat spacetime for the final fate of non-spherical collapse: 
The horizons of gravity form when and only when a mass $M$ 
gets compacted in a region whose circumference in {\it every} 
direction obeys $ {\cal C}\le 2\pi(2GM/c^2)$. Thus, unlike 
the cosmic censorship, the hoop conjecture does not rule out 
{\it all} naked singularities but only makes a definite assertion 
on the occurrence of event horizons in gravitational collapse. 
The hoop conjecture is concerned with the formation of event 
horizons, and not with naked singularities. Thus, even 
when event horizons form, say for example in the spherically 
symmetric case, it does not rule out the existence of naked 
singularities, or it does not imply that such horizons must
always cover the singularities.

When the collapse is sufficiently aspherical, with one 
or two dimensions being sufficiently larger than the others, 
the final state of collapse could be a naked singularity, 
according to the hoop conjecture. Such a situation
is inspired by the 
Lin, Mestel and Shu (1965)  
instability consideration in Newtonian gravity, 
where a non-rotating homogeneous spheroid collapses 
maintaining its homogeneity and spheroidicity but with
growing deformations. If the initial condition is that of 
a slightly oblate spheroid, the collapse results into a
pancake singularity through which the evolution could proceed. 
But for a slightly prolate spheroidal configuration, the 
matter collapses to a thin thread which results into a spindle
singularity. The gravitational potential and the tidal forces 
blow up as opposed to only density blowing up so it is a 
serious singularity. Even in the case of an oblate collapse, 
the passing of matter through the pancake causes prolateness
and subsequently a spindle singularity again results without the
formation of any horizon.

It is clear though, that the non-spherical collapse 
scenario is rather complex to understand, and a recourse to
the numerical simulations of evolving collapse models may 
greatly enhance our understanding on possible collapse final 
states in this case. In such a context, the numerical
calculations of 
Shapiro and Teukolsky (1991)
indicated conformity with the hoop conjecture.
They evolved collissionless gas spheroids in full general 
relativity which collapse in all cases to singularities. 
When the spheroid is sufficiently compact a black hole may 
form, but when the semimajor axis of the spheroid is sufficiently 
large, a spindle singularity forms without an apparent horizon 
forming. This gives rise to the possibility of
occurrence of naked singularities in collapse of finite
systems in asymptotically flat spacetimes which  
violate weak cosmic censorship but are in accordance 
with the hoop conjecture.

We note that the Kerr black hole is believed to be the 
unique stationary solution in Einstein gravity when the mass 
and rotation parameters are included. But it is to be noted 
that while the Schwarzschild black hole is the final
endstate of the homogeneous dust collapse, we have no interior 
solution for a rotating collapsing cloud. In other words,
an exterior Kerr geometry has no internal solution in general
relativity. We therefore do not really know the final fate
of a gravitational collapse with rotation. To understand 
the same, numerical simulations in full general relativity
will be of great value. There are many such numerical 
programs in the making currently to deal with this problem 
of modeling a rotating collapsing massive star. The idea here
is to include rotation in collapse and then to let the Einstein
equations evolve the collapse to see if the Kerr black hole
necessarily emerges as final state (see e.g.
Giacomazzo et al, 2011 
and references therein). 
and references therein. It is worth noting that numerical 
simulations in higher dimensions have also recently produced 
some very intriguing naked singularity formation scenarios
Lehner et al (2010).

We finally note that apart from such numerical simulations, 
certain analytic treatments of aspherical collapse are also
available. For example, the non-spherical Szekeres models for 
irrotational dust without any Killing vectors, generalizing the 
spherical Tolman-Bondi-Lemaitre collapse, were studied by
Joshi and Krolak (1986)
to deduce the existence of strong curvature naked
singularities. While this indicates that naked singularities 
are not necessarily confined to spherical symmetry only, 
it is to be noted that dynamical evolution of a 
non-spherical collapse still remains a largely uncharted 
territory.

\section{Are naked singularities stable and generic?}
Naked singularities may develop in gravitational collapse,
either spherical or otherwise. However, if they are not either
generic or stable in some suitable sense, then they may not
be necessarily physically relevant.
An important question then is the genericity and stability 
of naked singularities arising from regular initial data. Will the 
initial data subspace, which gives rise to naked singularity 
as end state of collapse, have a vanishing measure in a suitable sense? 
In that case, one would be able to reformulate more suitably 
the censorship hypothesis, based on a criterion that naked 
singularities could form in collapse but may not be generic.

We note here that the {\it genericity} and {\it stability} 
of the collapse outcomes, in terms of
black holes and naked singularities need to be understood 
carefully and in further detail. It is by and large well-accepted
now, that the general theory of relativity does allow and
gives rise to both black holes and naked singularities as final
fate of a continual gravitational collapse, evolving from a
regular initial data, and under reasonable physical conditions.
What is not fully clear as yet is the distribution
of these outcomes in the space of all allowed outcomes
of collapse. The collapse models discussed above and considerations
we gave here would be of some help in this direction, and may
throw some light on the distribution of black holes and
naked singularity solutions in the initial data space.
For some considerations on this issue, especially in the context
of scalar field collapse, we refer to
Christodoulou (1999), 
and
Joshi, Malafarina and Saraykar (2011),
and references therein, for further discussion.  
For the case of inhomogeneous dust collapse, the black hole
and naked singularity spaces are shown in 
Fig 4.

\begin{center}
\leavevmode\epsfysize=3.5 in\epsfbox{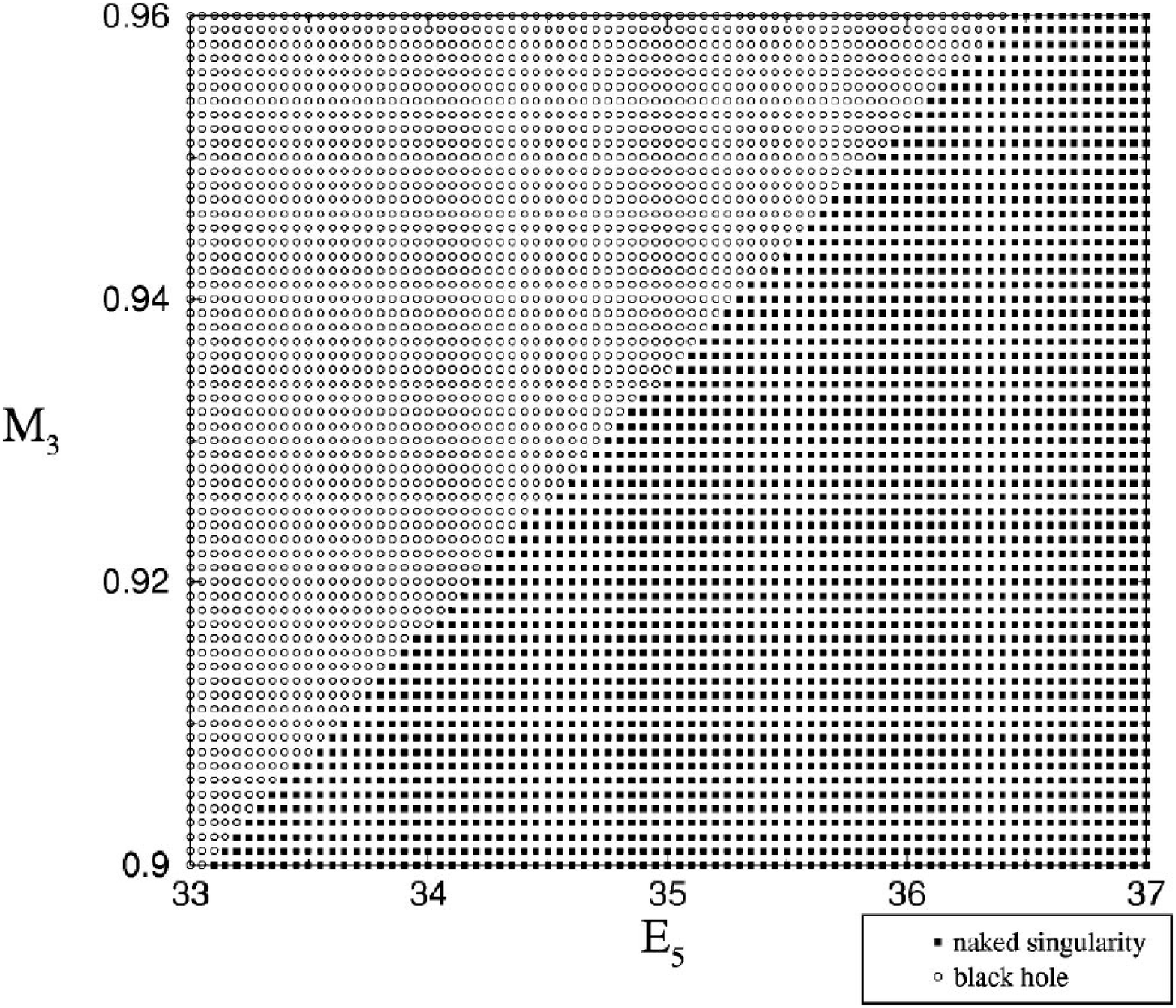}
\end{center}
\noindent {\small Fig 4: In the spaces of mass functions and 
energy functions, the initial data leading to black holes and
naked singularities are shown.}

The important point, however, is in general relativity 
there is no well defined concept or formulation as to what 
to call generic and stable outcomes, unlike the Newtonian case.
In other words, there are no well-defined criteria or 
definition available as to what is meant by stability in general
relativity. The ambiguity mainly arises because of non-unique 
topologies on the space of all Lorentzian metrics on a given
spacetime manifold, and a similar non-uniqueness of measures.
Under the situation, there is no easy way to answer this question
in any unique and definite manner, and people generally resort 
to the physical meaningfulness of the collapse scenario which 
gives rise to either the black hole or the naked singularity
outcome.

From such a perspective, it is natural and meaningful 
to ask here, what is really the physics that causes a naked 
singularity to develop in collapse, rather than a black hole? 
We need to know how at all
particles and energy are allowed to escape from extremely strong
gravity fields. We have examined this issue in some detail to bring
out the role of inhomogeneities and spacetime shear towards 
distorting the geometry of horizons that form in collapse
(see Fig 5).

\begin{center}
\leavevmode\epsfysize=3.5 in\epsfbox{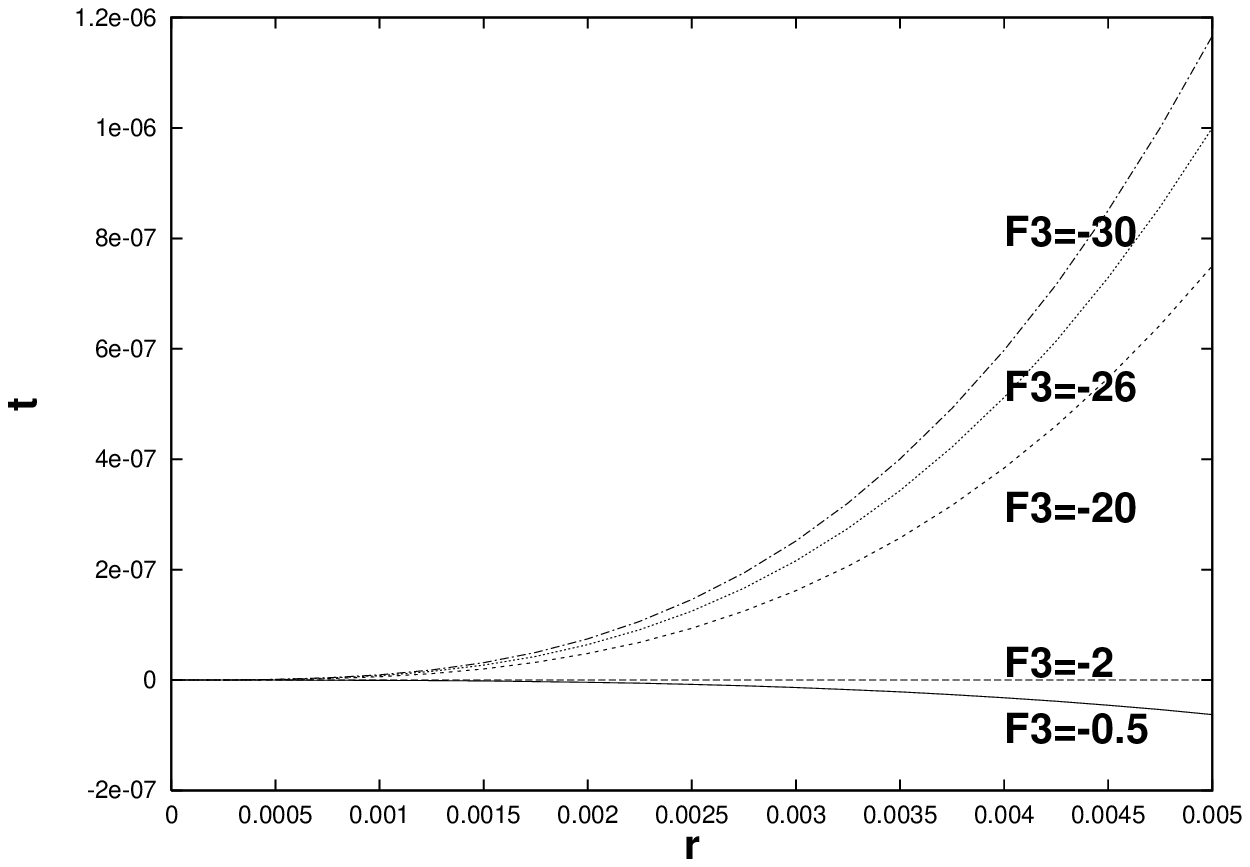}
\end{center}
\noindent {\small Fig 5: The apparent horizon formation is delayed
depending on the amount of inhomogeneity present as the collapse
proceeds (from Joshi, Dadhich and Maartens, 2002).}

In Newtonian gravity, it is only the matter density that determines the
gravitational field. In Einstein theory, however, density is only
one attribute of the overall gravitational field, and the various
curvature components and scalar quantities play an equally important
role to dictate what the overall nature of the field is.
What we showed is, once the density is inhomogeneous          
or higher at the center of
the collapsing star, this rather naturally delays the trapping of light
and matter during collapse, which can in principle escape away. This
is a general relativistic effect to imply that even if the densities 
are very high, there are paths available for light or matter to 
escape due to inhomogeneously collapsing matter fields.
These physical features then naturally lead to a naked singularity
formation
(Joshi, Dadhich and Maartens, 2002).

As it turns out, it is the amount of
inhomogeneity that counts towards distorting the apparent horizon
formation. 
If it is very small, below a critical limit, a black hole will 
form, but with sufficient inhomogeneity the trapping is delayed 
to cause a naked singularity. This criticality also comes out 
in the Vaidya class of radiation collapse models, where it 
is the rate of collapse, that is how fast or slow the cloud 
is collapsing, that determines the black hole or naked 
singularity formation.

\section{Astrophysical and observational aspects}
It is clear that the black hole and naked singularity
outcomes of a complete gravitational collapse for a massive
star are very different from each other physically, and
would have quite different observational signatures.
In the naked singularity case, if it occurs in nature,
we have the possibility to observe the physical effects
happening in the vicinity of the ultra dense regions that form
in the very final stages of collapse. However, in a black
hole scenario, such regions are necessarily hidden
within the event horizon of gravity.

There have been attempts where researchers explored 
physical applications and implications of the naked singularities 
(see e.g. 
Joshi and Malafarina 2011
and references in there).
If we could find astrophysical applications of the models
that predict naked singularities as collapse final fate, and possibly 
try to test the same through observational methods and the signatures 
predicted, that could offer a very interesting avenue to get 
further insight into the problem as a whole. An attractive recent 
possibility in that connection is to explore the naked singularities 
as possible particle accelerators 
(Patil and Joshi 2011),
where the possibility also emerges that the Cauchy Horizons may
not be innocuous, and high energy collisions could occur in the
vicinity of the same if they are generated by naked singularity
(see Fig 6).

\begin{center}
\leavevmode\epsfysize=3.5 in\epsfbox{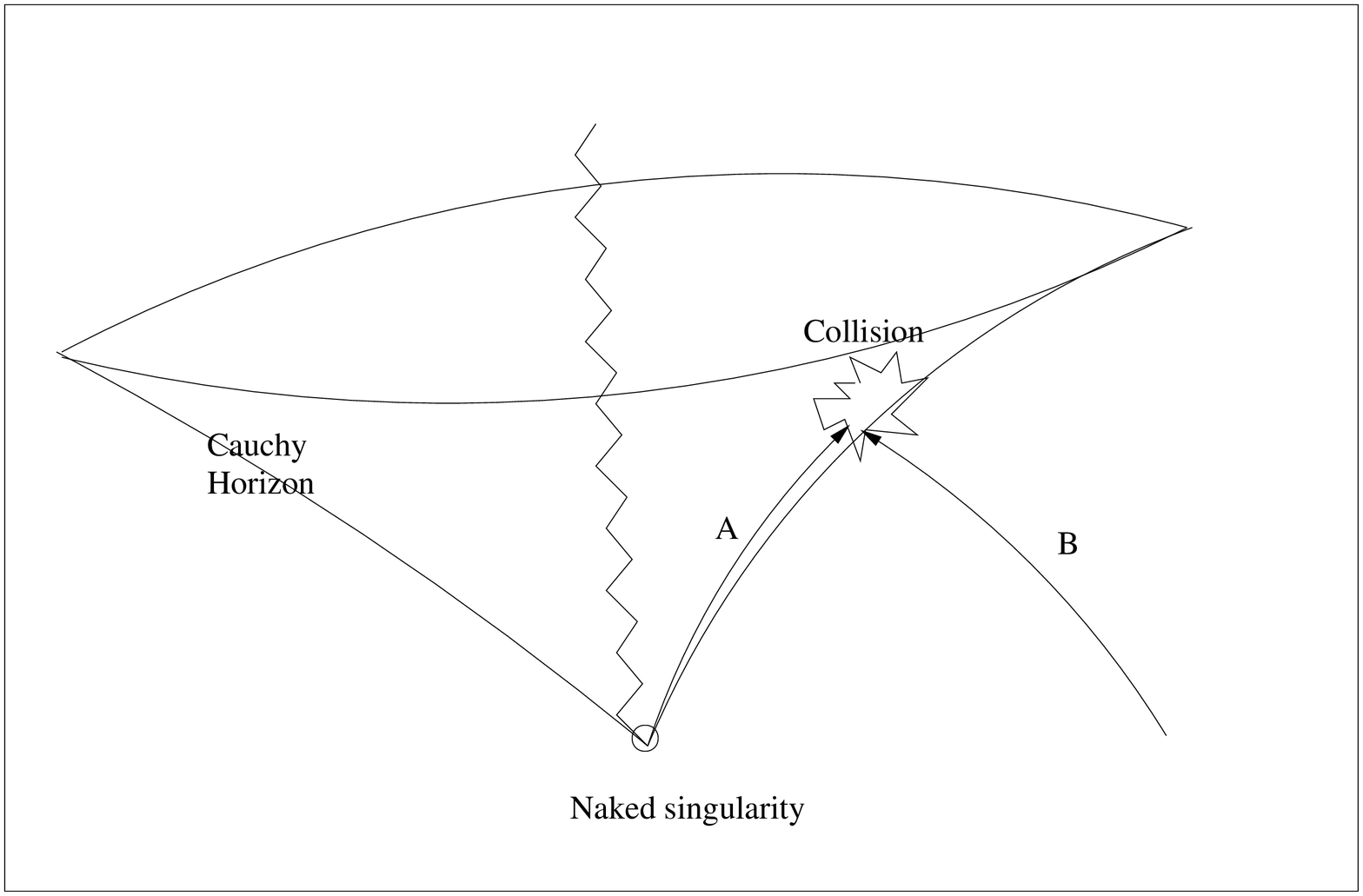}
\end{center}
\noindent {\small Fig 6: Very high energy particle collisions
can occur in the vicinity of the Cauchy horizon emerging from the 
naked singularity.}

Also, the accretion discs around a naked singularity,
wherein the matter particles are attracted towards or repulsed
away from the singularities with great velocities could provide
an excellent venue to test such effects and may lead to
predictions of important observational signatures to
distinguish the black holes and naked singularities in
astrophysical phenomena. The question of what observational signatures 
would then emerge and distinguish the black holes from naked singularities
is then necessary to be investigated, and we must explore
what special astrophysical consequences the latter may have.

One may ask several intriguing questions such as:{\it Where could the 
observational signatures of naked singularities lie?} 
If we look for the sign of singularities such as the ones that
appear at the end of collapse, we have to consider explosive and high
energy events. In fact such models expose the ultra-high density
region at the time of formation of the singularity while the outer
shells are still falling towards the center. In such a case,
shock-waves emanating from the superdense region at scales smaller than
the Schwarzschild radius (that could be due to quantum effects or
repulsive classical effects) and collisions of particles near the
Cauchy horizon could have effects on the outer layers. These would be
considerably different from those appearing during the formation of 
a black hole. If, on the other hand, we consider singularities such as the
super-spinning Kerr solution we can look for different kinds of
observational signatures. Among these the most prominent features deal
with the way the singularity could affect incoming particles, either
in the form of light bending, such as in gravitational lensing,
particle collisions close to the singularity, or properties of
accretion disks.

Essentially what we ask is: {\it Whether we could test censorship using 
astronomical observations}.
With so many high technology power missions to observe the
cosmos, can we not just observe the skies carefully to determine
the validity or otherwise of the cosmic censorship?
In this connection, several proposals to measure the mass and 
spin ratio for compact objects and for the galactic center 
have been made by different researchers. 
In particular, using pulsar observations it is suggested
that gravitational waves and the spectra of X-rays binaries 
could test the rotation parameter for the center of
our galaxy. Also, the shadow cast by the compact object 
can be used to test the same in stellar mass objects, 
or X-ray energy spectrum emitted by the
accretion disk can be used. Using certain observable properties 
of gravitational lensing that depend upon rotation
is also suggested (for references, see
Joshi and Malafarina, 2011).

The basic issue here is that of sensitivity, namely how
accurately and precisely can we measure and determine these
parameters. A number of present and future astronomical
missions could be of help. One of these is the Square-Kilometer
Array (SKA) radio telescope, which will offer a possibility
here, with a collecting area exceeding a factor of hundred
compared to existing ones. The SKA astronomers point out they
will have the sensitivity desired to measure the required quantities
very precisely to determine the vital fundamental issues in
gravitation physics such as the cosmic censorship, and
to decide on its validity or otherwise.
Other missions that could in principle provide a huge
amount of observational data are those that are currently hunting
for the gravitational waves. Gravitational wave astronomy
has yet to claim its first detection of waves, nevertheless
in the coming years it is very likely that the first
observations will be made by the experiments
such as LIGO and VIRGO that are currently still below
the threshold for observation. Then gravitational wave
astronomy will become an active field with possibly large
amounts of data to be checked against theoretical
predictions and it appears almost certain that this
will have a strong impact on open theoretical issues
such as the Cosmic Censorship problem.

There are three different kinds of observations that one could
devise in order to distinguish a naked singularity from a black hole.
The first one relies on the study of accretion disks. The accretion
properties of particles falling onto a naked singularity would be very
different from those of black hole of the same mass (see for example 
(Pugliese et al, 2011),
and the resulting accretion disks would also be observationally
different. The properties of accretion disks have been studied
in terms of the radiant energy, flux and luminosity, in a Kerr-like
geometry with a naked singularity, and the differences from a black 
hole accretion disk have been investigated.
Also, the presence of a naked singularity gives rise to powerful
repulsive forces that create an outflow of particles from the
accretion disk on the equatorial plane.
This outflow that is otherwise not present in the black hole case,
could be in principle distinguished from the jets of particles
that are thought to be ejected from black hole's polar region and
which are due to strong electromagnetic
fields. Also, when charged test particles are considered
the accretion disk's properties
for the naked singularity present in the Reissner-Nordstrom spacetime
are seen to be observationally different from those
of black holes.

The second way of distinguishing black holes from naked singularities relies
on gravitational lensing. It is argued that when the spacetime does not
possess a photon sphere, then the lensing features of light passing close to
the singularity will be observationally different from those of a black hole.
This method, however, does not appear to be very effective when a
photon sphere is present in the spacetime (see e.g.
Virbhadra et al 1998, 
and for recent update and further references,
Sahu et al, 2012, 2013, 
and
Joshi and Malafarina, 2011).  
Assuming that a Kerr-like solution of Einstein equations with
massless scalar field exists at the center of galaxies,
its lensing properties are studied
and it was found that there are effects due to the presence of
both the rotation and scalar field that would affect the behavior of
the bending angle of the light ray, thus making those objects
observationally different from black holes (see e.g.
Schee and Stuchlik, 2009).

Finally, a third way of distinguishing black holes from naked singularities
comes from particle collisions and particle acceleration in the vicinity of
the singularity. In fact, it is possible that the repulsive effects due to
the singularity can deviate a class of infalling particles, making
these outgoing eventually. These could then collide with some
ingoing particle, and the energy of collision
could be arbitrarily high, depending on the impact parameter of the
outgoing particle with respect to the singularity. The net effect is
thus creating a very high energy collision that resembles that of
an immense particle accelerator and that would be impossible in
the vicinity of a Kerr black hole.

It was pointed our recently by
Joshi, Malafarina and Ramesh Narayan (2011)
that one could obtain equilibrium configurations as final
outcome of a gravitational collapse. If such an object arises
without trapped surfaces but with a singularity at the center
(see Fig 7)
then again the accretion disk properties are very different 
from a black hole of the same mass.

\begin{center}
\leavevmode\epsfysize=3.5 in\epsfbox{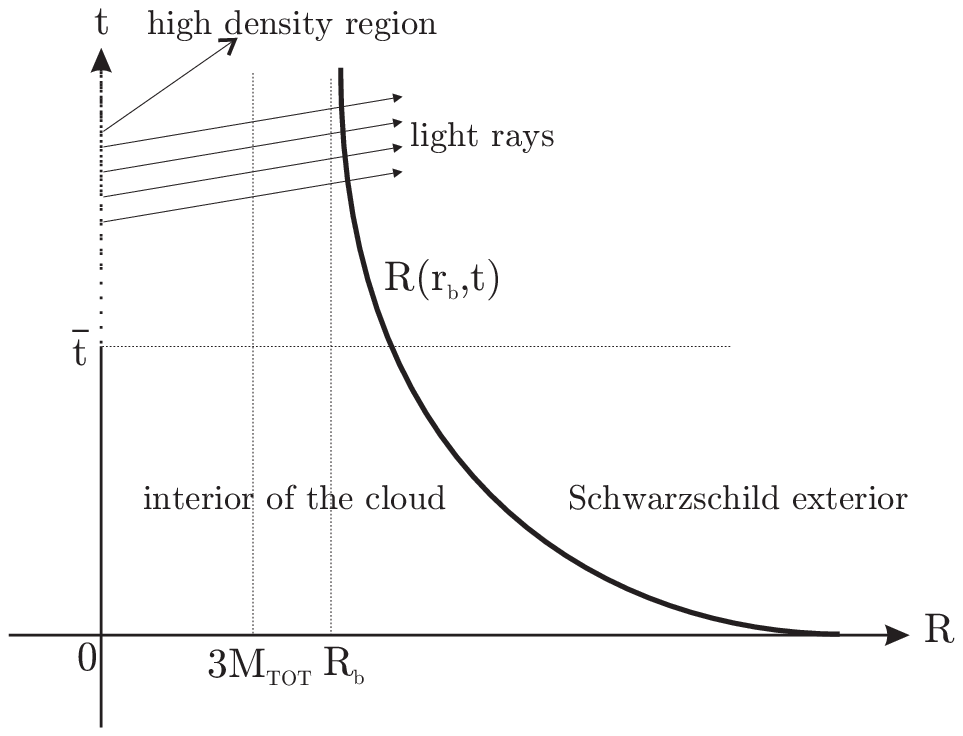}
\end{center}
\noindent {\small Fig 7: An equilibrium configuration can be
obtained from gravitational collapse which halts asymptotically,
and which would contain a central naked singularity.}

\section{Predictability and other cosmic puzzles}
What then is the status of naked singularities versus censorship 
today? Can cosmic censorship survive in some limited and specialized
form, and firstly, can we properly formulate it after all these 
studies in recent years on gravitational collapse?
While this continues to be a major cosmic puzzle, recent 
studies on formation of naked singularities as collapse end states 
for many realistic models have 
brought to forefront some of the most intriguing basic questions, both
at classical and quantum level, which may have significant physical
relevance. Some of these are: Can the super ultra-dense regions
forming in a physically realistic collapse of a massive star be
visible to far away observers in space-time? Are there any observable
astrophysical consequences? What is the causal structure of spacetime
in the vicinity of singularity as decided by the internal dynamics of
collapse which evolves from a regular initial data at an initial
time?  How early or late the horizons will actually develop in a
physically realistic gravitational collapse, as determined by the 
astrophysical conditions within the star? When a naked 
singularity forms, is it possible to observe the quantum gravity 
effects taking place in the ultra-strong gravity regions? Can one 
possibly envisage a connection to observed ultra-high energy 
phenomena such as cosmic gamma ray bursts?

A continuing study of collapse phenomena 
within a general and physically realistic framework may be the only
way to answers on some of these issues. This could lead us to 
novel physical insights and possibilities emerging out of the 
intricacies of gravitational force and nature of gravity, as emerging 
from examining the dynamical evolutions as allowed by 
Einstein equations.

Apart from its physical relevance, the collapse phenomena 
also have profound philosophical implications such as on the issue 
of predictability in the universe. We summarize below a few arguments, 
for and against it in the classical general relativity.

It is sometimes argued that breakdown of censorship means violation
of predictability in spacetime, because we have no direct handle 
to know what a naked singularity may radiate and emit unless we
study the physics in such ultra-dense regions. One would not be able
then to predict the universe in the future of a given epoch of time as
would be the case, for example, in the case of the Schwarzschild black
hole that develops in Oppenheimer-Snyder collapse.
A concern usually expressed is if naked
singularities occurred as the final fate of gravitational
collapse, predictability is violated in the spacetime, because 
the naked singularity is characterized by the existence of light 
rays and particles that emerge from the same. Typically, in all 
the collapse models discussed above, there is a family of future 
directed non-spacelike
curves that reach external observers, and when extended
in the past these meet the singularity.
The first light ray that comes out from the singularity
marks the boundary of the region that can be predicted
from a regular initial Cauchy surface in the spacetime,
and that is called the Cauchy horizon for the spacetime.
The causal structure of spacetime would differ
significantly in the two cases, when there is a Cauchy
horizon and when there is none.

In general relativity, a given `epoch' of time is sometimes          
represented by a spacelike surface, which is a three-dimensional 
space section. 
For example, in the standard Friedmann models of cosmology, there
is such an epoch of simultaneity, from which the universe evolves in
future, given the physical variables and initial data on this
surface. The Einstein equations govern this evolution of universe, and
there is thus a predictability which one would expect to hold in a
classical theory. The concern then is one would
not be able to predict in the future of naked singularity, and that
unpredictable inputs may emerge from the same.

Given a regular initial data on a
spacelike hypersurface, one would like to predict the future
and past evolutions in the spacetime for all times (see for example, 
Hawking and Ellis 1973).
Such a requirement is termed as the {\it global hyperbolicity}
of the spacetime. A globally hyperbolic spacetime is a fully
predictable universe, it admits a {\it Cauchy surface}, 
which is a three dimensional spacelike surface the data
on which can be evolved for all times in the past as well
as in future. Simple enough spacetimes such as the Minkowski
or Schwarzschild are globally hyperbolic, but the Reissner-Nordstrom
or Kerr geometries are not globally hyperbolic. For further
details on these issues, we refer to 
(Joshi, 2008).

The key role that the event horizon of a black hole plays
is that it hides the super-ultra-dense region formed in collapse
from us. So the fact that we do not understand such regions
has no effect on our ability to predict what happens
in the universe at large. But if no such horizon exists, then
the ultra-dense region might, in fact, play an important and
even decisive role in the rest of the universe, and our ignorance
of such regions would become of more than merely academic
interest.

Yet such an unpredictability is common in general relativity,
and not always directly related to censorship violation. Even black
holes themselves need not fully respect predictability when
they rotate or have some charge. For example, if we drop an
electric charge into an uncharged black hole, the spacetime geometry
radically changes and is no longer predictable from a regular
initial epoch of time. A charged black hole admits a naked
singularity which is visible to an observer within the horizon,
and similar situation holds when the black hole is rotating.
There is an important debate in recent years, if one could 
over-charge or over-rotate a black hole so that the singularity
visible to observers within the horizon becomes visible to external
far away observers too
(see e.g. Joshi 2009).

Also, if such a black hole was big enough on a
cosmological scale, the observer within the horizon could
survive in principle for millions of years happily without
actually falling into the singularity, and would thus be able
to observe the naked singularity for a long time. Thus,
only purest of pure black holes with no charge or rotation at
all respect the full predictability, and all other physically
realistic ones with charge or rotation actually do not.
As such, there are many models of the universe in cosmology
and relativity that are not totally predictable from a given
spacelike hypersurface in the past. In these universes, the
spacetime cannot be neatly separated into space and time foliation
so as to allow initial data at a given moment of time to
fully determine the future.

Actually the real breakdown of predictability is the
occurrence of spacetime singularity itself we could say, 
which indicates the true limitation of the classical gravity 
theory. It does not matter really whether it is hidden 
within an event horizon
or not. The real solution of the problem would then be the
resolution of singularity itself, through either a quantum
theory of gravity or in some way at the classical level.

In fact the cosmic censorship way to predictability,
that of `hiding the singularity within a black hole', and
then thinking that we restored the spacetime predictability
may not be the real solution, or at best it may be only a partial
solution to the key issue of predictability in spacetime universes.
It may be just shifting the problem elsewhere, and some
of the current major paradoxes faced by the black hole physics
such as the information paradox, the various puzzles regarding
the nature of the Hawking radiation, and other issues could
as well be a manifestation of the same.

No doubt, the biggest argument in support of censorship would 
be that it would justify and validate the extensive formalism and 
laws of black hole physics and its astrophysical applications 
made so far.  Censorship has been the foundation
for the laws of black holes such as the area theorem and others,
and their astrophysical applications. But these are not free
of major paradoxes.
Even if we accept that all massive stars 
would necessarily turn into black holes, this still creates 
some major physical paradoxes. Firstly, all the matter entering 
a black hole must of necessity collapse into a space-time singularity of
infinite density and curvatures, where all known laws of physics 
break down, which is some kind of instability at the classical 
level itself. This was a reason why many gravitation theorists 
of 1940s and 1950s objected to black hole formation, and Einstein 
also repeatedly argued against such a final fate of a collapsing star, 
writing a paper in 1939 to this effect.
Also, as is well-known and has been widely discussed in the 
past few years, a black hole, by potentially destroying information, 
appears to contradict the basic principles of quantum theory. In that
sense, the very formation of a black hole itself with a
singularity within it appears to come laden with inherent
problems. It is far from clear how one would resolve these
basic troubles even if censorship were correct.

In view of such problems with the black hole paradigm,
a possibility worth considering is the delay or avoidance of
horizon formation as the star collapses under gravity. This
happens when collapse to a naked singularity takes place, namely,
where the horizon does not form early enough or is avoided.
In such a case, if the star could radiate away most of its mass
in the late stages of collapse, this may offer a way out of
the black hole conundrum, while also resolving the singularity
issue, because now there is no mass left to form the singularity.
While this may be difficult to achieve purely classically,
such a phenomenon could happen when quantum gravity effects are
taken into account (Fig 8, see also the next section for
a further discussion).

\begin{center}
\leavevmode\epsfysize=3.5 in\epsfbox{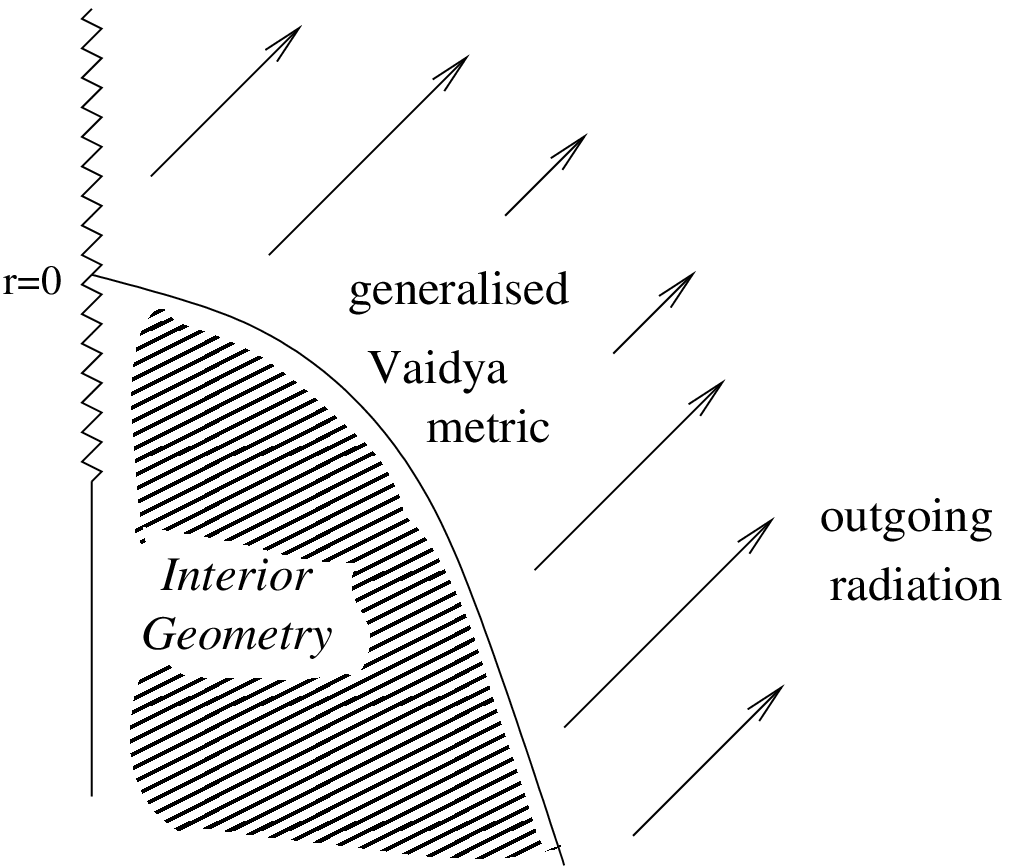}
\end{center}
\noindent {\small Fig 8: If the star could radiate away very
considerable mass, especially through negative quantum pressures 
close to the classical singularity, this may effectively resolve
the singularity.}

What this means is, such an `unpredictability' is somewhat
common in general relativity. For example, if we drop a slight charge 
in a Schwarzschild black hole, the spacetime geometry
completely changes into that of a charged black hole that is no longer
predictable in the above sense. Similar situation holds when the 
black hole is rotating. In fact, there are very many models of universe
in use in relativity which are not `globally hyperbolic', that is, not
totally predictable in the above sense where space and time are 
neatly separated so as to allow initial data to fully determine future
for all times.

In any case, a positive and useful feature that has emerged
from work on collapse models 
so far is, we already have now several important constraints 
for any possible formulation of censorship. It is seen that several 
versions of censorship proposed earlier would not hold,
because explicit counter-examples are available now.
Clearly, analyzing gravitational collapse
plays a crucial role here. Only if we understand clearly why naked
singularities do develop as collapse endstates in many realistic
models, there could emerge any pointer or lead to any practical 
and provable version of censorship.

Finally, it may be worth noting that even if the problem of singularity 
was resolved somehow,
possibly by invoking quantum gravity which may smear the singularity, 
we still have to mathematically formulate and prove the black 
hole formation assuming an appropriate censorship principle, which 
is turning out to be most difficult task with no sign of resolve. 
As discussed, the detailed collapse calculations of recent years 
show that the final fate of a collapsing star could be a naked
singularity in violation to censorship.  Finally, as is well-known 
and widely discussed by now, a black hole creates the information 
loss paradox, violating unitarity and making contradiction with 
basic principles of quantum theory. It is far from clear how one 
would resolve these basic troubles even if censorship were correct.

\section{A Lab for quantum gravity--Quantum stars?}
It is believed that when we have a reasonable and complete 
quantum theory of gravity available, all spacetime singularities, 
whether naked or those hidden inside black holes, will be 
resolved away. As of now, it remains an open question if the quantum
gravity will remove naked singularities.
After all, the occurrence of spacetime singularities
could be a purely classical phenomenon, and
whether they are naked or covered should not be relevant,
because quantum gravity will possibly remove them
all any way. 
It is possible that in a suitable quantum gravity theory
the singularities will be smeared out, though this has been
not realized so far.

In any case, the important and real issue is,
whether the extreme strong gravity regions formed due
to gravitational collapse are visible to faraway observers
or not. It is quite clear that the gravitational collapse
would certainly proceed classically, at least till the
quantum gravity starts governing and dominating the
dynamical evolution at the scales of the order
of the Planck length, {\it i.e.} till the extreme gravity
configurations have been already developed due to
collapse. The point is, it is the visibility or
otherwise of such ultra-dense regions that is under
discussion, whether they are classical or quantum
(see Fig 9).


\begin{center}
\leavevmode\epsfysize=3.5 in\epsfbox{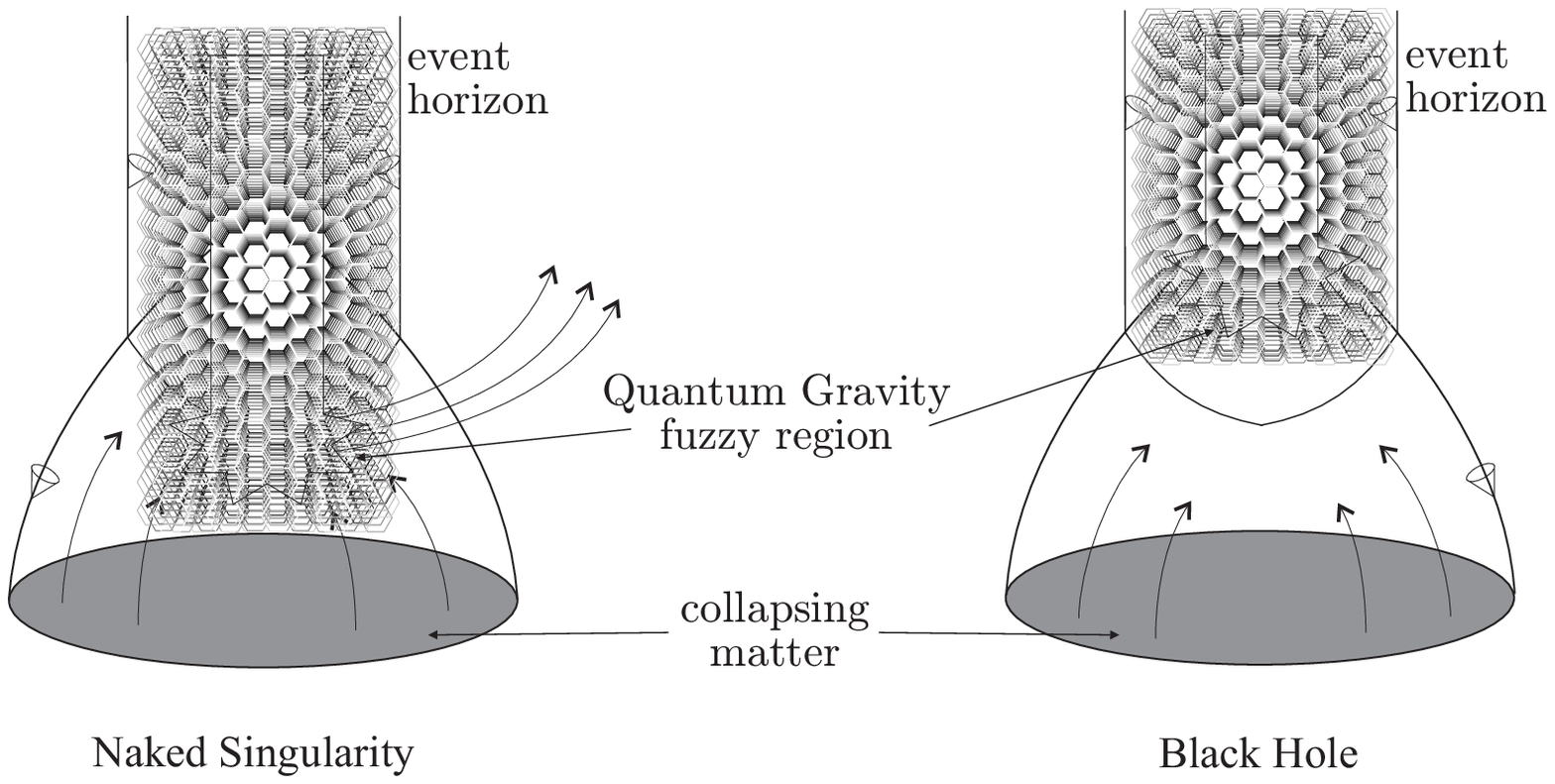}
\end{center}
\noindent {\small Fig 9: The naked singularity may be resolved by
quantum gravity effects but the ultra-strong
gravity region that developed in gravitational collapse
will still be visible to external observers
in the universe.}

What is important is, classical gravity implies
necessarily the existence of ultra-strong gravity
regions, where both classical and quantum gravity come into
their own. In fact, if naked singularities do develop in
gravitational collapse, then in a literal sense we come
face-to-face with the laws of quantum gravity, whenever
such an event occurs in the universe.

In this way, the gravitational collapse phenomenon
has the potential to provide us with a possibility of
actually testing the laws of quantum gravity.
In the case of a black hole developing in the
collapse of a finite sized object such as a massive star,
such strong gravity regions are necessarily hidden
behind an event horizon of gravity, and this would be
well before the physical conditions became extreme
near the spacetime singularity.
In that case, the quantum effects, even if they caused
qualitative changes closer to singularity, will be
of no physical consequences as no causal
communications are then allowed from such regions. On
the other hand, if the causal structure were that
of a naked singularity, then the communications from
such a quantum gravity dominated extreme curvature
ball would be visible in principle. This will be so
either through direct physical processes near a
strong curvature naked singularity, or via the
secondary effects, such as the shocks produced in
the surrounding medium.
It is possible that a spacetime singularity basically
represents the incompleteness of the classical theory and
when quantum effects are combined with the gravitational
force, the classical singularity may be resolved.

Therefore, more than the existence of a naked singularity,
the important physical issue is whether the extreme
gravity regions formed in the gravitational collapse of a
massive star are visible to external observers in the universe.
An affirmative answer here would mean that such a collapse
provides a good laboratory to study quantum gravity effects in
the cosmos, which may possibly generate clues for an as yet
unknown theory of quantum gravity. Quantum gravity theories
in the making, such as the string theory or loop quantum
gravity in fact are badly in need of some kind of an observational
input, without which it is nearly impossible to constrain
the plethora of possibilities.

We could say quite realistically that a laboratory similar
to that provided by the early universe is created in the collapse
of a massive star. However, the big bang, which is also
a naked singularity in that it is in principle visible to all
observers, happened only once in the life of the universe
and is therefore a unique event. But a naked singularity of
gravitational collapse could offer an opportunity to explore
and observe the quantum gravity effects every time a massive
star in the universe ends its life.

The important questions one could ask are: If in realistic
astrophysical situations the star terminates as a naked singularity,
would there be any observable consequences which reflect the
quantum gravity signatures in the ultra-strong gravity region?
Do naked singularities have physical properties different
from those of a black hole? Such questions underlie our
study of gravitational collapse.

In view of recent results on gravitational collapse, and 
various problems with the black hole paradigm, a
possibility worth considering is the delay or avoidance of
horizon formation as the star evolves collapsing under
gravity. This happens when collapse to a naked singularity
takes place, where the horizon does not form early enough or
is avoided.  In such a case, in the late stages of collapse
if the star could radiate away most of its mass,
then this may offer a way out of the black hole conundrum, while also
resolving the singularity issue, because now there is no mass left to
form the curvature singularity. The purpose is to resolve the black
hole paradoxes and avoid the singularity, either visible or within a
black hole, which actually indicates the breakdown of physical theory.
The current work on gravitational collapse suggests possibilities in
this direction.

In this context, we
considered a cloud that collapsed to a naked singularity
final state, and introduced loop quantum gravity effects
(Goswami, Joshi and Singh, 2006).
It turned out that the quantum effects generated an extremely
powerful repulsive force within the cloud. Classically the cloud would
have terminated into a naked singularity, but quantum effects
caused a burstlike emission of matter in the very last phases of
collapse, thus dispersing the star and dissolving the naked
singularity. The density remained finite and the spacetime
singularity was eventually avoided.  
One could expect this to
be a fundamental feature of other quantum gravity theories
as well, but more work would be required to confirm such a
conjecture.

For a realistic star, its final catastrophic collapse takes
place in matter of seconds. A star that lived millions of
years thus collapses in only tens of seconds. In the very last
fraction of a microsecond, almost a quarter of its total mass
must be emitted due to quantum effects, and therefore this
would appear like a massive, abrupt burst to an external observer
far away. Typically, such a burst will also carry with it specific
signatures of quantum effects taking place in such ultra-dense
regions. In our case, these included a sudden dip in the
intensity of emission just before the final burstlike evaporation
due to quantum gravity. The question is, whether such unique 
astrophysical signatures
can be detected by modern experiments, and if so, what they tell
on quantum gravity, and if there are any new insights into
other aspects of cosmology and fundamental theories such as
string theory.
The key point is, because the very final ultra-dense regions
of the star are no longer hidden within a horizon as in the black
hole case, the exciting possibility of observing these quantum
effects arises now, independently of the quantum gravity theory
used. An astrophysical connection to extreme high energy
phenomena in the universe, such as the gamma-rays bursts that defy
any explanations so far, may not be ruled out.

Such a resolution of naked singularity through quantum gravity
could be a solution to some of the paradoxes mentioned above. 
Then, whenever
a massive star undergoes a gravitational collapse, this might
create a laboratory for quantum gravity in the form of a
{\it Quantum Star}
(see e.g. Joshi, 2009),
that we may be able to possibly access.
This would also suggest intriguing connections to high
energy astrophysical phenomena. The present situation poses
one of the most interesting
challenges which have emerged through the recent work on
gravitational collapse.

We hope the considerations here have shown that gravitational
collapse, which essentially is the investigation of dynamical
evolutions of matter fields
under the force of gravity in the spacetime,
provides one of the most exciting research frontiers in gravitation
physics and high energy astrophysics.
In our view, there is a scope therefore for both theoretical
as well as numerical investigations in these areas, which
may have much to tell for our quest on basic issues in quantum
gravity, fundamental physics and gravity theories, and towards
the expanding frontiers of modern high energy astrophysical
observations.

\section{Concluding remarks}
We considered here several aspects of spacetime singularities
and the physical scenarios where these may be relevant, playing 
an interesting and intriguing role. We hope this creates a 
fairly good view of the exciting new physics that the spacetime 
singularities are leading us to, presenting a whole spectrum of
new possibilities towards our search of the universe.

After discussing their existence and certain key basic properties,
we discussed in some detail the gravitational collapse scenarios 
and the useful conclusions that have emerged so far in this context. 
In the first place, singularities not covered fully by the
event horizon do occur in several collapsing configurations from 
regular initial data, with reasonable equations of state such as 
describing radiation, dust or a perfect fluid with a non-zero
pressure, or also for general forms of matter. These naked singularities 
are physically significant in that densities and curvatures diverge powerfully
near the same. Such results on the final fate of collapse, generated 
from study of different physically reasonable collapse scenarios, 
may provide useful insights into black hole physics and may be of help 
for any possible formulation of the cosmic censorship hypothesis.

An insight that seems to emerge is, the final state of a collapsing 
star, in terms of either a black hole or a naked singularity, may not 
really depend on the form or equation of state of collapsing matter, 
but is actually determined by the physical initial data in terms 
of the initial density profiles and pressures.

As an example, for inhomogeneous dust collapse, the final fate
could be a black hole or a naked singularity depending on the values 
of initial parameters. The collapse ends in a naked singularity
if the leading nonvanishing derivative of density at the center is 
either the first one or the second one. There is a transition from 
the naked singularity phase to the black hole phase as the initial 
density profile is made more and more homogeneous near the center. 
As one progresses towards more homogeneity, and hence towards
a stronger gravitational field, there first occurs a weak naked 
singularity, then a strong naked singularity, and finally 
a black hole.

The important question then is the genericity and stability 
of such naked singularities arising from regular initial data. 
Will the initial data subspace giving rise to naked singularity 
have zero measure in a suitable sense? In that case, one would be 
able to reformulate more suitably the censorship hypothesis, 
based on a criterion that naked singularities could form in 
collapse but may not be generic. As we pointed out, the answer 
is far from clear due to ambiguities in the definitions of
measures and the stability criteria. 

One may try to evolve some kind of a physical formulation 
for cosmic censorship, where the available studies on various 
gravitational collapse scenarios such as above may provide a
useful guide. The various properties of naked singularities
collectively may be studied as they emerge from the studies so far
and one would then argue that objects with such properties are 
not physical. But the way forward is again far from clear.

One could also invoke quantum effects and quantum gravity. 
While naked singularities may form in classical general relativity, 
quantum gravity presumably removes them. The point is that even though 
the final singularity may be removed in this way, still there would 
be very high density and curvature regions in the classical regime
which would be causally communicating with outside observers, as 
opposed to the black hole case. If quantum effects could remove the 
naked singularity, this would then be some kind of {\it quantum 
cosmic censorship}.

We hope the considerations here have shown that gravitational
collapse, which essentially is the investigation of dynamical
evolutions of matter fields under the force of gravity in the spacetime,
provides one of the most exciting research frontiers in gravitation
physics and high energy astrophysics.
There are issues here which have deep relevance both for
theory as well as observational aspects in astrophysics and
cosmology. Also these problems are of relevance
for basics of gravitation theory and quantum gravity, and these
inspire a philosophical interest and inquiry into the
nature and structure of spacetime, causality, and profound
issues such as predictability in the universe, as we
indicated here.

Research is already happening in many of these
areas as the discussion here pointed out. Some of the most
interesting questions from my personal perspective are: 
Genericity and stability of collapse outcomes, examining the 
quantum gravity effects near singularities, observational 
and astrophysical signatures of the collapse outcomes, 
and other related issues. In particular, one of the most interesting 
questions would be, if naked singularities which are hypothetical 
astrophysical objects, did actually form in nature, what distinct 
observational signatures they would present. That is, how one 
distinguishes the black holes from naked singularities would 
be an important issue. There have been some efforts on this issue 
in recent years as we indicated above.
The point is, there are already very high energy astrophysical
phenomena being observed today, with several observational 
missions working both from ground and space. The black holes and naked
singularities which are logical consequences of star collapse in
general relativity, would appear to be the leading candidates to
explain these phenomena. The observational signatures
that each of these would present, and their astrophysical
consequences would be of much interest for the future
theoretical and computational research, and for their
astrophysical applications.

In our view, there is therefore a scope for both theoretical
as well as numerical investigations in these frontier areas, which
may have much to tell for our quest on basic issues in quantum
gravity, fundamental physics and gravity theories, and towards
the expanding frontiers of modern high energy astrophysical
observations.

\bigskip

{\bf References}
\bigskip

Choptuik M. W. (1993) , 'Universality and scaling in gravitational 
collapse of a massless scalar field', Phys. Rev. Lett. 70, p.9.

Christodoulou, D. (1999), 'The instability of naked singularities 
in the gravitational collapse of a scalar field', Ann. of Maths. {\bf
149}, 183.

Clarke C. J. S. (1975), `Singularities in globally hyperbolic
spacetimes', Commun.Math.Phys. {\bf 41}, 65.

Clarke C. J. S.  (1986),  `Singularities: Global and Local 
Aspects', in `Topological Properties and Global Structure of 
Space-time' (eds.  P. G. Bergmann and V.  de Sabbata),  Plenum 
Press,  New York.

Clarke C. J. S. and de Felice F. (1982), `Globally non-causal spacetimes',
J. Phys, {bf A15}, 2415.

Clarke C. J. S. and Joshi P. S. (1988), `On reflecting spacetimes', Class.Quantum
grav., {\bf 5}, 19.

Clarke C.  J.  S. and Krolak A.  (1986),  `Conditions for the 
Occurrence of Strong Curvature Singularities',    J.   Geo.   
Phys.   2,   p.  127.

Datt B., (1938), `$\ddot U$ber eineklasse von l$\ddot o$sungen der 
gravitationsgleichungen der relativit$\ddot a$t', Z. Physik, {\bf 
108}, 314.

Ellis G. F. R. and King A. (1974), `Was the big bang a whimper?',
Commun.Math.Phys. {\bf 38}, 119.

Ellis G. F. R. and Schmidt, B. (1977), `Singular spacetimes',
Gen.Relat.Grav. {\bf 8}, 915.

Giacomazzo B., Rezzolla L. and Stergioulas N. (2011),
`Collapse of differentially rotating neutron stars and 
cosmic censorship', Phys.Rev. {\bf D84}, 024022.

G\"odel K.  (1949),  `An Example of a New type of Cosmological 
Solution of Einstein's field Equations of Gravitation',  Rev. Mod. 
Phys., 21,  p. 447.

Goswami R. and Joshi P. S. (2007), `Spherical gravitational 
collapse in N-dimensions', Phys.Rev. {\bf D76}, 084026.

Goswami R., Joshi, P. S. and Singh P. (2006), 'Quantum 
Evaporation of a Naked Singularity'  Phys. Rev. Lett., {\bf96}, 
031302.

Hawking S. W. and Ellis G. F. R., (1973), {\it The large scale 
structure of spacetime}, Cambridge Univ. Press, Cambridge.

Hawking S.  W.   and Penrose R.  (1970),  `The Singularities of  
Gravitational Collapse and Cosmology', Proc.  R.  Soc.  Lond.   
A314,   529.

Joshi P. S. (1981), `On Higher Order Causality Violations', Phys. 
Lett. 85A, 319.

Joshi P. S. (1993),{\it Global aspects in gravitation and 
cosmology}, Oxford Univ. Press, Oxford.

Joshi P. S. (2008),{\it Gravitational collapse and spacetime 
singularities},
Cambridge Univ. Press, Cambridge.

Joshi P. S. (2009), {\it Naked Singularities}, Scientific 
American, {\bf 300}, 36.

Joshi P. S., Dadhich N., and Maartens R. (2002), `Why do naked 
singularities form in gravitational collapse?', Phys.Rev. {\bf D65},  
101501.

Joshi P. S. and Dwivedi I.  H.  (1992), `The Structure of Naked 
Singularity in Self-similar Gravitational Collapse', Commun. Math. 
Phys. {\bf 146},  333.

Joshi P. S. and Dwivedi I. H. (1993), `Naked singularities in 
spherically symmetric inhomogeneous Tolman-Bondi dust cloud collapse', 
Phys. Rev. {\bf D47}, 5357.

Joshi P. S. and Dwivedi I. H. (1999), `Initial data and 
the end state of spherically symmetric gravitational collapse',
Class.Quant.Grav. {\bf 16}, 41.

Joshi P. S. and Krolak A. (1996), 'Naked strong curvature 
singularities in Szekeres spacetimes', Class.Quant.Grav.{\bf 13}, 
3069.

Joshi P. S. and Malafarina D. (2011), `Recent developments 
in gravitational collapse and spacetime singularities', 
Int.J.Mod.Phys. {\bf D20}, 2641.

Joshi P. S., Malafarina D. and Ramesh Narayan (2011), 
`Equilibrium configurations from gravitational collapse', Class.
Quant.Grav. {\bf 28}, 235018.

Joshi P. S., Malafarina D. and Saraykar R. V. (2012), 
`Genericity aspects in gravitational collapse to black holes 
and naked singularities, Int.J.Mod.Phys. {\bf D21}, 1250066.

Joshi P. S.  and Saraykar R. V. (1987),  `Cosmic Censorship and 
Topology Change in General Relativity',  Phys. Lett.  {\bf 120A},  
111.

Kriele M. (1990), `Causality violations and singularities', 
Gen.Rel.Grav. {\bf 22}, 619.

Lake K. (1992),'Precursory singularities in spherical 
gravitational collapse', Phys. Rev. Lett. {\bf 68}, p.3129.

Lehner L. and Pretorius F. (2010), `Black Strings, Low Viscosity 
Fluids, and Violation of Cosmic 
Censorship', Phys.Rev.Lett. {\bf 105}, 101102.

Lin C. C.,  Mestel L.  and Shu F. H.  (1965), `The Gravitational 
Collapse of a Uniform Spheroid',  Astrophys.  J.,  142,  p. 1431.

Oppenheimer J.  R.   and Snyder H.  (1939),  `On Continued  
Gravitational Contraction',   Phys.  Rev.   56,   455.

Ori A.   and Piran T.  (1987),  `Naked Singularity in self-similar 
Spherical Gravitational Collapse', Phys. Rev. Lett. 59, p. 2137.

Patil M. and and Joshi P. S. (2011), `Kerr Naked Singularities as 
Particle Accelerators', Class.Quant.Grav. {\bf28}, 235012.

Penrose R.  (1979), `Singularities and Time asymmetry', in 
`General Relativity-an Einstein Centenary Survey' (ed. S. W. 
Hawking and W. Israel), Cambridge University, Press, Cambridge.

Pugliese D., Quevedo H., Ruffini R. (2011),
`Motion of charged test particles in Reissner-Nordstrom spacetime'
Phys.Rev. {\bf D83}, 104052.

Raychaudhuri A. K. (1955), `Relativistic Cosmology',  Phys.  Rev., 
98,  p. 1123.

Sahu S., Patil M., Narasimha D., Joshi P. S. (2012),
`Can strong gravitational lensing distinguish naked singularities 
from black holes?', Phys.Rev. D86, 063010, 
arXiv:1206.3077 [gr-qc]

Sahu S., Patil M., Narasimha D., Joshi P. S. (2013),
`Time Delay between Relativistic Images as a probe of Cosmic
Censorship', arXiv:1310.5350 [gr-qc], To appear in Phys.Rev.D.

Schee J. and Stuchlik Z. (2009),
`Optical phenomena in the field of braneworld Kerr black holes',
Int.J.Mod.Phys. D18, 983.

Schoen R.  and Yau S. -T.  (1983), `The Existence of a Black Hole 
due to condensation of Matter',  Commun. Math. Phys.   90, p. 575.

Senovilla J. M. M (1990), `New class of inhomogeneous cosmological 
perfect-fluid solutions without big-bang singularity', - 
Phys.Rev.Lett. {\bf 64}, 2219.

Shapiro S. L. and Teukolsky S. A. (1991), 
`Formation of naked singularities: The violation of cosmic 
censorship', Phys.Rev.Lett. {\bf 66}, 994.

Szekeres P. and Iyer V. (1993), 'Spherically symmetric 
singularities and strong cosmic censorship', Phys. Rev. {\bf D47}, 
p.4362.

Thorne K.  S. (1972),  `Non-spherical Gravitational Collapse: A 
Short Review' in `Magic without Magic$-$John Archibald Wheeler', 
ed.  J.  Clauder,  W.  H.  Freeman,   New York.

Tipler F. (1976), `Causality violations in asymptotically flat
spacetimes', Phys.Rev.Letts. {\bf 37}, 879.

Tipler F. (1977), 'Singularities in conformally flat spacetimes', 
Phys. Lett.{\bf 64A} 8 (1977).

Tipler F.,   Clarke C.  J.  S.   and Ellis G.  F.  R.  (1980),
`Singularities and Horizons' in `General relativity and 
Gravitation'  Vol 2,   (ed. A Held) Plenum,   New York.

Virbhadra K. S., Narasimha, D., Chitre S. M. (1998),
`Role of the scalar field in gravitational lensing',
Astron.Astrophys. 337, 1, e-Print: astro-ph/9801174.

Wald R. {\it General Relativity}, U.of Chicago Press, Chicago 
(1984).

\end{document}